\documentclass[lettersize,journal]{IEEEtran}
\usepackage[caption=false,font=normalsize,labelfont=sf,textfont=sf]{subfig}
\usepackage{stfloats}
\usepackage{verbatim}
\def\BibTeX{{\rm B\kern-.05em{\sc i\kern-.025em b}\kern-.08em
    T\kern-.1667em\lower.7ex\hbox{E}\kern-.125emX}}
\usepackage{balance}

\usepackage[numbers,square,sort&compress]{natbib}
\usepackage{graphicx}
\usepackage{amsmath,amssymb,amsfonts}
\usepackage{amsthm}
\usepackage{algorithmic}
\usepackage{textcomp}
\usepackage{float}
\usepackage{etoolbox}
\usepackage{hyperref}
\usepackage{orcidlink}
\usepackage{tikz}
\usetikzlibrary{arrows.meta}
\usepackage{fontawesome}
\usepackage{subcaption}
\usepackage{url}
\usepackage{listings}
\usepackage{pdfpages}
\usepackage[english]{babel}
\usepackage{scalerel}
\usepackage{xcolor,colortbl}
\usepackage[linesnumbered,ruled,vlined]{algorithm2e}
\usepackage{amsmath}
\usepackage{amssymb}
\usepackage{multirow}
\usepackage{array}
\usepackage{lipsum}
\usepackage{booktabs} 
\usepackage{tabularx} 

\newtheorem{lemma}{Lemma}

\theoremstyle{definition}

\newtheorem{theorem}{Theorem}

\providecommand{\sample}{\hskip2.3pt{\gets\!\!\mbox{\tiny${\$}$\normalsize}}\,}

\newcommand{\PartySecretKey}[1]{\ensuremath{\textrm{sk}_{#1}}}
\newcommand{\Party}[1]{\ensuremath{P_{#1}}}
\newcommand{\Parties}{\ensuremath{\mathbb{P}}}
\newcommand{\Corrupted}{\ensuremath{\mathcal{C}}}

\newcommand{\PublicKey}{\textbf{E}}
\newcommand{\SecretKey}{\textbf{d}}

\newcommand{\PartialSecretKey}[1]{\ensuremath{d_{#1}}}
\newcommand{\PartialPublicKey}[1]{\ensuremath{E_{#1}}}

\newcommand{\SharePartialSecretKey}[2]{\ensuremath{[d_{#1}]_{#2}}}

\newcommand{\EncryptedSharePartialSecretKey}[2]{\ensuremath{C_{#1,#2}}}

\newcommand{\SetOfFDKG}{\ensuremath{\mathbb{D}}}

\newcommand{\Voters}{\ensuremath{\mathbb{V}}}
\newcommand{\Tallies}{\ensuremath{\mathbb{T}}}

\newcommand{\Ballot}[1]{\ensuremath{B_{#1}}}

\newcommand{\BlindingFactor}[1]{\ensuremath{r_{i}}}
\newcommand{\Vote}[1]{\ensuremath{v_{#1}}}

\newcommand{\GuardianSetOf}[1]{\ensuremath{\mathcal{G}_{#1}}}
\newcommand{\TotalA}{\ensuremath{C1}}
\newcommand{\TotalB}{\ensuremath{C2}}
\newcommand{\BallotA}[1]{\ensuremath{C1_{#1}}}
\newcommand{\BallotB}[1]{\ensuremath{C2_{#1}}}

\newcommand{\G}{\ensuremath{G}}

\newcommand{\SharePartialDecryptionFromTo}[2]{\ensuremath{[\mathrm{PD}_{#1}]_{#2}}}

\newcommand{\PartialDecryptionFrom}[1]{\ensuremath{\mathrm{PD}_{#1}}}

\begin{document}

\title{Federated Distributed Key Generation}


\author{Stanislaw Baranski\,\orcidlink{0000-0001-7181-8860}     
    \and
    Julian Szymanski\,\orcidlink{0000-0001-5029-6768}
    \thanks{Stanislaw Baranski and Julian Szymanski are with Department of Electronic, Telecommunication and Informatics, Gdansk University of Technology, Gdansk, 
Poland}
}

\date{}
\maketitle

\begin{abstract}
    Distributed Key Generation (DKG) underpins threshold cryptography across many domains: decentralized custody and wallets, validator/committee key ceremonies, cross-chain bridges, threshold signatures, secure multiparty computation, and internet voting.

    Classical $(t,n)$-DKG assumes a fixed set of $n$ parties and a global threshold $t$, and effectively requires full, timely participation; when actual participation or availability deviates, setups abort or must be rerun—a severe problem in open or ad-hoc, time-critical settings, especially when $n$ is large and node availability is unpredictable.”

    We introduce \emph{Federated Distributed Key Generation (FDKG)}, inspired by Federated Byzantine Agreement, that makes participation optional and trust \emph{heterogeneous}. Each participant selects a personal guardian set $\GuardianSetOf{i}$ of size $k$ and a local threshold $t$; its partial secret can later be reconstructed either by itself or by any $t$ of its guardians. FDKG generalizes PVSS-based DKG and completes both \emph{generation} and \emph{reconstruction} in a single broadcast round each, with total communication $\Theta(n\ k)$ and (in the worst case) $\mathcal{O}(n^2)$ for reconstruction.

    Our security analysis shows that: (i) \emph{Generation} achieves correctness, privacy, and robustness under standard PVSS-based DKG assumptions; and (ii) \emph{Reconstruction} provides liveness and privacy characterized by the \emph{guardian-set topology} $\{\GuardianSetOf{i}\}$. Liveness holds provided that, for every participant $i$, the adversary does not control $i$ together with at least $k - t + 1$ of its guardians. Conversely, reconstruction privacy holds unless the corrupted set itself is reconstruction-capable.

    In a scenario with $n{=}100$ potential parties ($|\SetOfFDKG|{=}50$, $k{=}40$, 80\% retention for tallying), FDKG distribution broadcasts $\approx 336$\,kB and reconstruction $\approx 525$\,kB. Client-side Groth16 proving per participant is $\approx 5$\,s for distribution, and $0.6$–$29.6$\,s for reconstruction depending on the number of revealed shares.
\end{abstract}

\section{Introduction}
\label{sec:introduction}

Distributed Key Generation (DKG) is a multi-party protocol that lets a group of mutually distrustful parties jointly create a public/secret key pair and hold it via threshold shares: any $t$ out of $n$ parties can reconstruct or use the secret, while fewer learn nothing \cite{pedersenThresholdCryptosystemTrusted1991, gennaroSecureDistributedKey1999}. By shifting key creation from a single entity to a committee, DKG removes single point of failure for core operations such as decryption or signing. DKGs are widely used across threshold encryption and signatures~\cite{gennaroSecureDistributedKey1999,pedersenThresholdCryptosystemTrusted1991,shoupSecuringThresholdCryptosystems1998,boldyrevaThresholdSignaturesMultisignatures2002,kokoris-kogiasCALYPSOPrivateData2018}, consensus protocols~\cite{cachinRandomOraclesConstantinople2000}, multi party computation (MPC)~\cite{cramerMultipartyComputationThreshold2000,damgardUniversallyComposableEfficient2003,brakerskiThresholdFHEEfficient2025}, distributed randomness beacons (DRB) \cite{choiSoKDistributedRandomness2023, dasDistributedRandomnessUsing2024,kavousiSoKPublicRandomness2023}, cryptocurrency wallets~\cite{gennaroThresholdoptimalDSAECDSA2016,LITProtocolWhitepaper2024} and internet voting protocols~\cite{adidaHeliosWebbasedOpenAudit2008, cortierBeleniosSimplePrivate2019,haenniCHVoteProtocolSpecification2017,baranskiTrustCentricApproachQuantifying2024}.

However, a standard $(t,n)$-DKG protocol \cite{pedersenThresholdCryptosystemTrusted1991,gennaroSecureDistributedKey1999} requires $n$ to be known a priori to determine polynomial degrees and sharing parameters. In settings where participation is voluntary (e.g., public blockchains or ad hoc networks), the realized participant count $n'$ may only become clear after an initial interaction phase. If $n'$ deviates from the anticipated $n$, the pre-selected global threshold $t$ may be unattainable ($t>n'$) or misaligned with the actual corruption/availability assumptions; in either case the system must be reparameterized and restarted, incurring coordination cost and delay that are often unacceptable in time-critical deployments. Furthermore, \textbf{traditional DKGs enforce uniform (homogeneous) trust among all $n$ parties, discarding the organic trust structure that could otherwise be leveraged to improve security and availability.}

To overcome these limitations, we introduce \emph{Federated Distributed Key Generation (FDKG)}, a protocol that draws conceptual inspiration from Federated Byzantine Agreement (FBA) protocols~\cite{mazieresStellarConsensusProtocol2015}, designed for scenarios with uncertain participation and heterogeneous trust relationships. FDKG allows any subset of potential parties to participate in the key generation process. Its core innovation lies in \emph{Guardian Sets}: each participating party $\Party{i}$ independently selects a trusted subset of $k$ other parties (its guardians) and defines a local threshold $t$ necessary for these guardians to reconstruct $\Party{i}$'s partial secret key.

Formally, an FDKG protocol involves $n$ potential parties. A subset $\SetOfFDKG \subseteq \Parties$ chooses to participate in generating partial secret keys. Each participant $\Party{i} \in \SetOfFDKG$ distributes its partial secret key $\PartialSecretKey{i}$ among its chosen guardian set $\GuardianSetOf{i}$ of size $k$ using a local $(t,k)$-threshold secret sharing scheme. The global secret key $\SecretKey$ is the sum of all valid partial secrets $\PartialSecretKey{i}$. During reconstruction, $\SecretKey$ can be recovered if, for each $\Party{i} \in \SetOfFDKG$, either $\Party{i}$ itself provides $\PartialSecretKey{i}$, or at least $t$ of its guardians in $\GuardianSetOf{i}$ provide their respective shares. FDKG generalizes traditional $(t,n)$-DKG, which can be viewed as a special case where all $n$ parties participate and each party's guardian set comprises all other $n-1$ parties.

\begin{figure}
    \centering
    \includegraphics[width=0.5\textwidth]{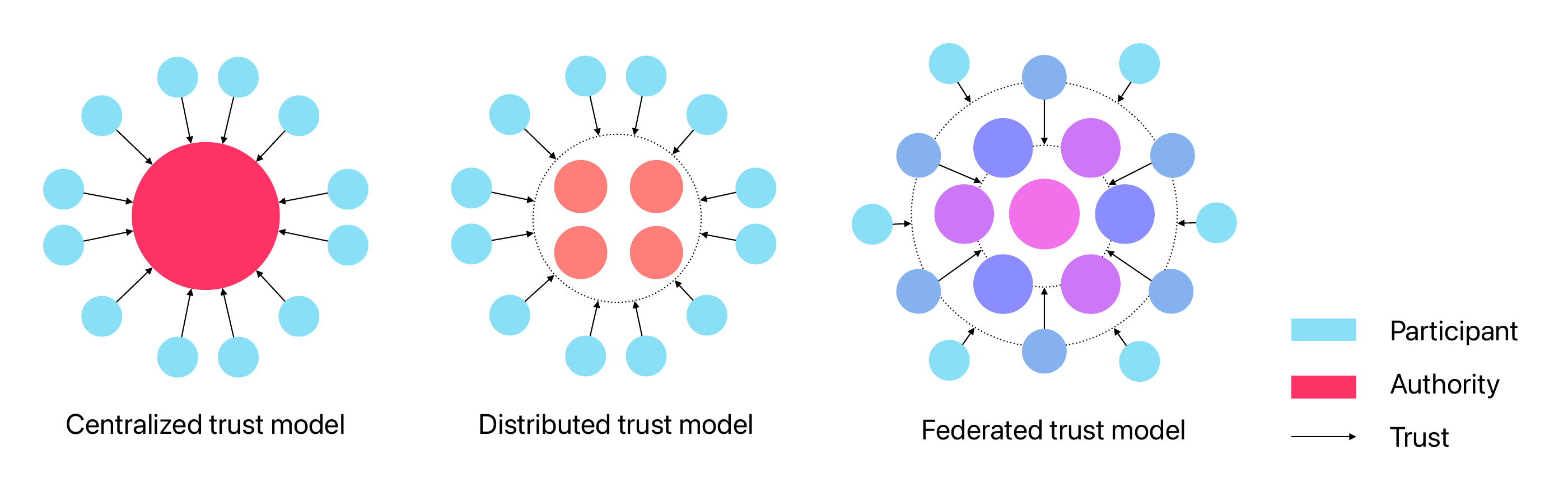}
    \caption{Comparison of trust models in distributed systems: centralized trust (one authority), distributed trust (multiple authorities), and federated trust (peer-to-peer with individually chosen trusted groups).}
    \label{fig:trust-models}
\end{figure}

This novel approach offers several advantages:
\begin{itemize}
    \item \textbf{Optional Participation:} FDKG accommodates any subset of parties $\Party{i} \in \SetOfFDKG$ joining the key generation process without a prior commitment phase, improving UX by reducing interaction and avoiding costly reparameterizations/restarts when $n' \neq n$.
    \item \textbf{Heterogeneous Trust Model:} By allowing each participant to define its own guardian set $\GuardianSetOf{i}$, FDKG supports diverse and evolving trust relationships, making it suitable for decentralized settings where trust is not uniform. This is illustrated in Figure~\ref{fig:trust-models}, which contrasts FDKG's federated trust with centralized and traditional distributed trust models.
    \item \textbf{Resilience to Node Unavailability:} The use of local thresholding via guardian sets ensures that a participant's contribution to the global key can be recovered even if the participant itself becomes unavailable, provided a sufficient number of its designated guardians remain honest and active.
\end{itemize}

In practice, FDKG succeeds in settings where classical $(t,n)$\mbox{-}DKG cannot tolerate participation uncertainty without costly restarts and timeline slips. Consider a public voting with a 24-hour enrollment window for talliers. The organizer publishes a list of $n=60$ eligible talliers and plans a classical DKG with $t=40$. After the window, only $n'=37$ complete enrollment and verification. Since classical DKG pre-fixes $t$ against the planned $n$, \textbf{the chosen $t=40$ is unattainable ($t>n'$), so setup must abort or be restarted with new parameters}. In FDKG, the same upper bound $n=60$ is published, but Round~1 accepts any subset $\SetOfFDKG$ of those who post valid transcripts (here $|\SetOfFDKG|=37$).

A time-critical disclosure scenario similarly benefits. A whistleblower wants to share files with up to $n=12$ independent journalists; the case is urgent, so he sets a 6-hour enrollment window. His goal is to include as many as possible by the deadline, encrypt once the window closes, and avoid being blocked by a few non-responders. With FDKG, the joint key emerges from the actual participants $\SetOfFDKG$, so progress does not hinge on every individual being present. Classical DKG, by contrast, requires full participation; if fewer than the planned $n$ attend within 6 hours, the setup fails and must be rerun—missing the deadline the whistleblower cannot extend.

In this paper, we provide a detailed specification of the FDKG protocol, a formal security analysis within the simulation-based paradigm (Section~\ref{sec:security_analysis}), and demonstrate its practical utility through an application to an internet voting scheme. Our primary contribution is the FDKG framework itself, which enables robust and verifiable key generation with adaptable trust for dynamic networks.

This paper makes the following contributions:
\begin{itemize}
    \item We formally define the FDKG protocol.
    \item We provide a security analysis in the simulation-based framework, establishing Correctness, Privacy, and Robustness for key generation, and Liveness for reconstruction (Section~\ref{sec:security_analysis}).
    \item We demonstrate FDKG's performance and resilience through extensive simulations under various network conditions, participation rates, and retention scenarios (Section~\ref{sec:liveness_simulations}), and evaluate the computational and communication costs of its cryptographic components (Section~\ref{sec:performance_evaluation}).
    \item We integrate FDKG into an end-to-end internet voting scheme, showcasing its practical application using threshold ElGamal encryption and NIZK-based verifiability (Section~\ref{sec:voting_scheme}). A corresponding open-source implementation, together with reproducibility scripts, is provided.
\end{itemize}

The remainder of this paper is organized as follows. Section~\ref{sec:related_work} reviews related work in DKG. Section~\ref{sec:preliminaries} presents the cryptographic primitives. Section~\ref{sec:fdkg} describes the FDKG protocol. Section~\ref{sec:security_analysis} provides the security analysis. Section~\ref{sec:liveness_simulations} discusses simulation results. Section~\ref{sec:voting_scheme} details the voting application. Section~\ref{sec:performance_evaluation} assesses performance. Section~\ref{sec:deployments} outlines deployment strategies. Finally, Section~\ref{sec:limitations-and-future} discusses limitations and future work, and Section~\ref{sec:discussion-conclusion} offers concluding remarks.

\begin{table*}[htbp]
\centering
\caption{Comparison of synchronous, biasable DKG protocols. Core columns follow Bacho–Kavousi~\cite{bachoSoKDlogbasedDistributed2025}. We split ``Corrupt'' into \emph{Privacy} and \emph{Liveness}, and add \emph{Trust} and \emph{Setup}. 
Privacy: adversary cannot learn the shared secret. 
Liveness: adversary cannot prevent honest parties from recovering the shared secret.
Let $\Corrupted \subseteq \Parties$ be the statically corrupted set, and let $R$ be the family of \emph{reconstruction-capable sets} (as in Eq.~\eqref{eq:R}).
\textbf{Communication} is total broadcast cost; $BC_n(b)$ is the cost for one party to broadcast $b$ bits to $n$ parties; $k$ denotes the guardian set size used in the FDKG.
\textbf{Rounds} denotes the number of broadcast rounds. 
\textbf{Sharing} indicates the secret-sharing type (VSS, PVSS, or APVSS). 
\textbf{Trust Model}: ``Homo'' (homogeneous) means a global, uniform trust assumption; ``Hetero'' (heterogeneous) allows participant-defined, non-uniform trust. 
\textbf{Setup}: ``Required'' means a fixed participant set must take part; ``Optional'' discovers participants from a larger pool.}
\label{tab:dkg_comparison_final}
\renewcommand{\arraystretch}{1.25}
\begin{tabularx}{\textwidth}{l l l l l l l l l}
\toprule
\textbf{Protocol} & \textbf{Year} & \textbf{Privacy} & \textbf{Liveness} & \textbf{Communication} & \textbf{Rounds} & \textbf{Sharing} & \textbf{Trust} & \textbf{Setup} \\
\midrule
Pedersen~\cite{pedersenThresholdCryptosystemTrusted1991} & 1991 & $|\Corrupted| < t < n/2$ & $|\Corrupted| < n - t$ & $n \ BC_n(\lambda n)$ & $3 \cdot BC_n$ & VSS  & Homo & Required \\
Fouque–Stern~\cite{fouqueOneThresholdDiscreteLog2001}    & 2001 & $|\Corrupted| < t < n$   & $|\Corrupted| < n - t$ & $n \ BC_n(\lambda n)$ & $1 \cdot BC_n$ & PVSS & Homo & Required \\
Kate et al.~\cite{kateConstantSizeCommitmentsPolynomials2010}   & 2010 & $|\Corrupted| < t < n/2$ & $|\Corrupted| < n - t$ & $n \ BC_n(\lambda + \delta\lambda n)$ & $3 \cdot BC_n$ & VSS  & Homo & Required \\
Gürkan et al.~\cite{gurkanAggregatableDistributedKey2021} & 2021 & $|\Corrupted| < t < \log n$ & $|\Corrupted| < n - t$ & $n\ BC_n(\lambda) + \ell^2BC_n(\lambda n)$ & $\ell \cdot BC_n$ & APVSS & Homo & Required \\
Groth~\cite{grothNoninteractiveDistributedKey2021} & 2021 & $|\Corrupted| < t < n$ & $|\Corrupted| < n - t$ & $n \ BC_n(\lambda^2 n)$ & $1 \cdot BC_n$ & PVSS & Homo & Required \\
Kate et al.~\cite{kateNoninteractiveVSSUsing2023}        & 2023 & $|\Corrupted| < t < n$   & $|\Corrupted| < n - t$ & $n \ BC_n(\lambda n)$ & $1 \cdot BC_n$ & PVSS & Homo & Required \\
Cascudo–David~\cite{cascudoPubliclyVerifiableSecret2023} & 2023 & $|\Corrupted| < t < n$ & $|\Corrupted| < n - t$ & $n \ BC_n(\lambda n)$ & $1 \cdot BC_n$ & PVSS & Homo & Required \\
Feng et al.~\cite{fengScalableAdaptivelySecure2023}        & 2023 & $|\Corrupted| < t < n/2$ & $|\Corrupted| < n - t$ & $\kappa \ BC_n(\lambda n) + n\lambda \kappa$ & $2 \cdot BC_n$ & VSS  & Homo & Required \\
Bacho et al.~\cite{bachoGRandLineAdaptivelySecure2023}   & 2023 & $|\Corrupted| < t < n/2$ & $|\Corrupted| < n - t$ & $BC_n(\lambda n) + n^2 \lambda \log n$ & $4 \cdot BC_n$ & APVSS & Homo & Required \\
Feng et al.~\cite{fengDragonDecentralizationCost2024}    & 2024 & $|\Corrupted| < t < n/2$ & $|\Corrupted| < n - t$ & $\sqrt{n}\ BC_n(\lambda n\ \kappa)$ & $2 \cdot BC_n$ & PVSS & Homo & Required \\
\midrule
\rowcolor{gray!20}
\textbf{FDKG (this work)} & \textbf{2024} &
$\boldsymbol{\mathcal{C} \notin R}$ &
$\boldsymbol{\exists\, S \in R:\ S \subseteq \Parties \setminus \mathcal{C}}$ &
$\boldsymbol{n \ BC_n(\lambda k)}$ &
$\boldsymbol{1 \cdot BC_n}$ &
\textbf{PVSS} & \textbf{Hetero} & \textbf{Optional} \\
\bottomrule
\end{tabularx}
\end{table*}

\section{Related Work}
\label{sec:related_work}

Foundational DKG protocols, such as~\cite{pedersenThresholdCryptosystemTrusted1991, gennaroSecureDistributedKey1999}, established the core paradigm: $n$ participants, each acting as a dealer for a partial secret, combine their contributions to form a single joint public key and distributed secret shares. These protocols are typically based on Verifiable Secret Sharing (VSS)~\cite{feldmanPracticalSchemeNoninteractive1987}, which allows participants to verify the integrity of their received shares.

Subsequent research has extensively explored DKG, focusing on various aspects such as robustness against adversarial behavior, efficiency, round complexity, security, and applicability in different network models (synchronous vs. asynchronous)~\cite{bachoSoKDlogbasedDistributed2025}. Considerable effort has been dedicated to enhancing DKG for reliable operation in challenging environments like the internet, with a focus on asynchronous settings and high thresholds of corruptions \cite{kateDistributedKeyGeneration2012, dasPracticalAsynchronousDistributed2022, dasPracticalAsynchronousHighthreshold2022, zhangPracticalAsynchronousDistributed2023}.  A significant body of work focuses on Publicly Verifiable Secret Sharing (PVSS) \cite{stadlerPubliclyVerifiableSecret1996,schoenmakersSimplePubliclyVerifiable1999}, where the correctness of shared information can be verified by any observer, not just the share recipients. PVSS is crucial for non-interactive DKG protocols, where parties broadcast their contributions along with proofs, minimizing interaction rounds. Groth's work on DKG and key resharing \cite{grothNoninteractiveDistributedKey2021} presents efficient constructions leveraging pairing-based cryptography and NIZK proofs, aiming for minimal rounds and public verifiability, which is a common goal for modern DKG deployed on transparent ledgers or bulletin boards.

The recent SoK by Bacho and Kavousi~\cite{bachoSoKDlogbasedDistributed2025} identifies two primary axes for classifying DKG protocols: the underlying network model and the secret sharing mechanism employed. To accurately position FDKG, we adopt this framework and establish a coherent baseline for comparison.

First, DKG protocols are distinguished by their network model assumption. \textbf{Synchronous protocols} assume bounded message delays, allowing them to operate in well-defined rounds. In contrast, \textbf{asynchronous protocols}~\cite{huDyCAPSAsynchronousDynamiccommittee2022} are designed to tolerate unbounded message delays, a significantly stronger guarantee that requires more complex consensus mechanisms and often results in higher communication overhead. As FDKG is designed for synchronous environments, we omit asynchronous DKG protocols from our direct comparison to ensure an evaluation of protocols operating under similar network guarantees.

A second critical distinction is the protocol's resilience to key biasing by a rushing adversary. Modern research differentiates between \textbf{biasable} protocols and \textbf{unbiasable} (or "fully secure") protocols~\cite{katzOptimalFullySecure2023}. Fully secure DKGs guarantee that the final public key's distribution is uniform and cannot be influenced by a malicious minority. This robustness typically requires additional rounds of communication (e.g., commit-reveal schemes). FDKG, in line with many practical protocols, prioritizes simplicity and a minimal round count, and is therefore biasable. Consequently, to evaluate FDKG against schemes with similar design trade-offs, we focus our comparison on the class of biasable protocols.

Furthermore, it is important to distinguish DKG from Dynamic and Proactive Secret Sharing (DPSS) schemes~\cite{maramCHURPDynamicCommitteeProactive2019}. DPSS protocols are designed for the long-term lifecycle management of a secret, enabling committees to evolve and shares to be proactively refreshed to defend against adaptive adversaries. Since FDKG focuses on the initial \textit{bootstrapping} of a key, DPSS protocols address a different problem domain and are thus not directly included in our setup-focused analysis.

Our design draws conceptual inspiration from Federated Byzantine Agreement (FBA) protocols—such as the Stellar Consensus Protocol \cite{mazieresStellarConsensusProtocol2015}—where nodes declare local ``quorum slices'' (trusted peers), enabling open participation without a globally pre-defined validator set. FDKG borrows this form of heterogeneous, participant-defined trust: each party $i$ specifies a guardian set $G_i$ with local threshold $t$, and a party included in many guardian sets becomes operationally more critical for liveness. The semantics, however, differ fundamentally. In FBA, safety and liveness are defined via quorum intersection and concern \emph{consensus} over a ledger; in FDKG, properties are defined via the reconstruction family $R$ (defined in Section~\ref{sec:security_analysis}) and concern a one-shot cryptographic reconstruction predicate. Thus, while both systems use local trust slices, FDKG makes no consensus claims; our guarantees (Theorems~\ref{thm:recon_privacy}–\ref{thm:recon_liveness}) address reconstruction privacy and liveness rather than agreement.

To establish a coherent baseline and facilitate a sound comparison, we therefore narrow our focus to the class of \textbf{synchronous, biasable DKG protocols}. Within this class, different secret sharing mechanisms are used, as detailed in the~\cite{bachoSoKDlogbasedDistributed2025}. These include complaint-based VSS, non-interactive Publicly Verifiable Secret Sharing (PVSS)~\cite{stadlerPubliclyVerifiableSecret1996a, fouqueOneThresholdDiscreteLog2001} (enabling one-round designs), and Aggregatable PVSS (APVSS)~\cite{gurkanAggregatableDistributedKey2021, bachoAdaptivelySecureAggregatable2023} (allowing transcript aggregation). Beyond the sharing mechanism, protocols differ in three comparison-relevant dimensions that we report in Table~\ref{tab:dkg_comparison_final}: (i) the \emph{privacy security} and \emph{liveness tolerance} for corrupted parties; (ii) the round complexity, ranging from \emph{one} broadcast round (PVSS) up to \emph{four} rounds (multi-phase APVSS/complaint flows); and (iii) the \emph{communication overhead}.

Our comparative analysis, presented in Table~\ref{tab:dkg_comparison_final}, evaluates FDKG against this established baseline. The table reveals a consistent architectural paradigm across all surveyed protocols. To the best of our knowledge, FDKG is the first protocol in its class to break from this paradigm in two fundamental ways. First, it introduces a \textbf{heterogeneous trust model}, moving away from the assumption that trust is global and uniform (``Homo''). Second, it supports \textbf{optional participation}, whereas all prior schemes require the participation of a fixed, pre-defined set of members (``Required''). These architectural changes are what enable FDKG to provide key generation in time-critical and ad-hoc environments.

\section{Preliminaries}
\label{sec:preliminaries}

This section introduces the cryptographic primitives, concepts, and notations foundational to our Federated Distributed Key Generation (FDKG) protocol.

\subsection{Notation}
Let $\lambda$ denote the computational security parameter. We work within a cyclic group $\mathbb{G}$ of prime order $q$ generated by $G \in \mathbb{G}$. Group operations are written multiplicatively, and scalar multiplication is denoted by $G^a$ for $a \in \mathbb{Z}_q$. We assume the Discrete Logarithm Problem (DLP) is hard in $\mathbb{G}$. All algorithms are assumed to be Probabilistic Polynomial-Time (PPT) unless stated otherwise. Sampling a random element $x$ from a set $S$ is denoted by $x \sample S$. We denote the set of potential parties as $\Parties = \{\Party{1}, \dots, \Party{n}\}$.

\subsection{Shamir's Secret Sharing (SSS)}
\label{subsec:sss}
A $(t, n)$-threshold Shamir Secret Sharing scheme over the finite field $\mathbb{Z}_q$ (where $q$ is prime and $q > n$) allows a dealer to share a secret $s \in \mathbb{Z}_q$ among $n$ parties $\Party{1}, \dots, \Party{n}$ such that any subset of $t$ or more parties can reconstruct $s$, while any subset of fewer than $t$ parties learns no information about $s$.
\begin{itemize}
    \item $(s_1, \dots, s_n) \leftarrow \textrm{Share}(s, t, n)$: The dealer chooses a random polynomial $f(X) = \sum_{k=0}^{t-1} a_k X^k \in \mathbb{Z}_q[X]$ of degree $t-1$ such that $f(0) = s$. For each party $\Party{i}$ ($i \in [n]$), the dealer computes the share $s_i = f(i)$.
    \item $s  \leftarrow \textrm{Reconstruct}(I, \{s_i\}_{i \in I})$: Given a set $I \subseteq [n]$ of at least $t$ indices and their corresponding shares $\{s_i\}_{i \in I}$, this algorithm computes $s = \sum_{i \in I} \lambda_i s_i$, where $\lambda_i$ are the Lagrange coefficients for the set $I$ evaluated at 0.
\end{itemize}
SSS provides perfect privacy against adversaries controlling fewer than $t$ parties.

\subsection{Verifiable and Publicly Verifiable Secret Sharing (VSS/PVSS)}
\label{subsec:vss_pvss}
VSS schemes enhance SSS by allowing parties to verify the consistency of their shares relative to some public information. PVSS strengthens this by enabling \textit{any} entity to perform this verification using only public data \cite{stadlerPubliclyVerifiableSecret1996, schoenmakersSimplePubliclyVerifiable1999}.
PVSS schemes typically involve the dealer encrypting each share $s_i$ under the recipient $\Party{i}$'s public key $\textrm{pk}_i$ and publishing these ciphertexts $C_i = \textrm{Enc}(\textrm{pk}_i, s_i)$ along with public commitments to the secret polynomial (e.g., $\{A_k = G^{a_k}\}$) and a public proof $\pi$ of consistency. Our FDKG protocol employs a PVSS scheme where each participant acts as a dealer.

\textbf{DKG from PVSS:} In a typical PVSS-based DKG \cite{katzOptimalFullySecure2023}, each party $\Party{k}$ acts as a dealer for a partial secret $d_k=f_k(0)$. $\Party{k}$ uses PVSS to distribute shares $d_{k,j} = f_k(j)$ to all parties $\Party{j}$, publishing commitments $\{E_{k,l} = G^{a_{k,l}}\}$, encrypted shares $\{C_{k,j}\}$, and a proof $\pi_k$. The final secret key is $d = \sum_k d_k$, the public key is $E = \prod_k E_{k,0}$, and party $\Party{i}$'s share is $d_i = \sum_k d_{k,i}$.

\subsection{Public Key Encryption (PKE)}
\label{subsec:pke}
A public-key encryption scheme $\mathcal{E} = (\textrm{KeyGen}, \textrm{Enc}, \textrm{Dec})$ consists of three algorithms:
\begin{itemize}
    \item $(\textrm{pk}_i, \textrm{sk}_i) \sample \textrm{KeyGen}(1^\lambda)$: Generates a public key $\textrm{pk}_i$ and a secret key $\textrm{sk}_i$.
    \item $c \leftarrow \textrm{Enc}(\textrm{pk}_i, m, \textrm{rand})$: Encrypts message $m$ under $\textrm{pk}_i$ using provided randomness $\textrm{rand}$ to produce ciphertext $c$.
    \item $m/\perp \leftarrow \textrm{Dec}(\textrm{sk}_i, c)$: Decrypts ciphertext $c$ using secret key $\textrm{sk}_i$ to recover message $m$ or outputs failure $\perp$.
\end{itemize}
We require \textbf{IND-CPA} security (Indistinguishable under Chosen Plaintext Attack).

\subsection{Non-Interactive Zero-Knowledge (NIZK) Proofs}
\label{subsec:nizk}
NIZK proofs allow proving the truth of an NP statement $x \in \mathcal{L}$ (for relation $\mathcal{R}$) without interaction and without revealing the witness $w$.

NIZKs require a setup phase. Different models exist regarding the nature and trust assumptions of this setup:
\begin{itemize}
    \item \textbf{Structured Reference String (SRS) / Trusted Setup:} Systems like Groth16 \cite{grothSizePairingBasedNoninteractive2016} require a CRS generated with specific mathematical structure. This structure is essential for proof succinctness and efficiency but typically requires a trusted setup ceremony (or an MPC equivalent) to generate the CRS without revealing a trapdoor. The CRS is specific to a particular circuit/relation.
    \item \textbf{Universal / Updatable Setup:} Systems like PLONK \cite{gabizonPLONKPermutationsLagrangebases2019} or Marlin \cite{chiesaMarlinPreprocessingZkSNARKs2019} use a CRS that can be used for any circuit up to a certain size (universal) and may allow the CRS to be updated securely without repeating a full trusted ceremony (updatable). They still require an initial setup phase, often trusted.
    \item \textbf{Transparent Setup / Random Oracle Model (ROM):} Systems like STARKs \cite{ben-sassonScalableTransparentPostquantum2018} require minimal setup assumptions, often relying only on public randomness (transparent) or operating in the idealized Random Oracle Model (ROM), where a public hash function is modeled as a truly random function. These avoid complex or trusted setup ceremonies.
\end{itemize}
Our current implementation utilizes Groth16 due to its efficiency and widespread tooling, thus operating in the CRS model and requiring a trusted setup. However, the FDKG framework itself is compatible with other NIZK systems.

A NIZK system (e.g., in the CRS model) provides:
\begin{itemize}
    \item $\textrm{crs} \sample \textrm{Setup}(1^\lambda, \mathcal{R})$: Generates the public CRS.
    \item $\pi \sample \textrm{Prove}(\textrm{crs}, x, w)$: Produces proof $\pi$.
    \item $b \leftarrow \textrm{Verify}(\textrm{crs}, x, \pi)$: Outputs $1$ (accept) or $0$ (reject).
\end{itemize}
We require standard Completeness, computational Soundness, and computational Zero-Knowledge.

\subsection{Encrypted Channels for Share Distribution}
\label{subsec:encrypted_channels}
Protocols like PVSS and DKG often require confidential transmission of secret shares ($s_i$) to specific recipients ($\Party{i}$) over a public broadcast channel. This is achieved using an IND-CPA secure PKE scheme. Each potential recipient $\Party{i}$ has a key pair $(\textrm{pk}_i, \textrm{sk}_i)$ where $\textrm{pk}_i$ is public. To send $s_i$ to $\Party{i}$, the sender computes $c = \textrm{Enc}(\textrm{pk}_i, s_i, \textrm{rand})$ and broadcasts $c$. Only $\Party{i}$ can compute $s_i = \textrm{Dec}(\textrm{sk}_i, c)$.

For circuit efficiency with Groth16, we use an ElGamal variant over BabyJub \cite{weijiekohElGamalEncryptionDecryption2020, jieWeijiekohElgamalbabyjub2023, steffenZeestarPrivateSmart2022}. The specific ElGamal variant encryption and decryption algorithms used in our implementation are detailed in Algorithm\ref{alg:encryption} and Algorithm\ref{alg:decryption}, respectively.

\begin{algorithm}[h!]
    \caption{ElGamal Encryption $\textrm{Enc}(\textrm{pk}_i, m, [k, r])$}
    \label{alg:encryption} 
    \textbf{Input:} Recipient public key $\textrm{pk}_i$, scalar message $m \in \mathbb{Z}_q$, randomness $(k, r) \in \mathbb{Z}_q \times \mathbb{Z}_q$\\
    \textbf{Output:} Ciphertext tuple $(C_1, C_2, \Delta)$\\
    $C_1 = G^k$ \hspace{14.9em} \\
    $M = G^r$ \hspace{15.4em} \\
    $C_2 = (\textrm{pk}_i)^k \cdot M$ \hspace{10.2em}  \\
    $\Delta = M.x - m \pmod{q}$ \hspace{1.5em} \\
    \textbf{return} $(C_1, C_2, \Delta)$
\end{algorithm}

\begin{algorithm}[h!]
    \caption{ElGamal Decryption $\textrm{Dec}(\textrm{sk}_i, (C_1, C_2, \Delta))$}
    \label{alg:decryption} 
    \textbf{Input:} Recipient secret key $\textrm{sk}_i$, ciphertext $(C_1, C_2, \Delta)$\\
    \textbf{Output:} Scalar message $m \in \mathbb{Z}_q$\\
    $M = C_2 \cdot (C_1^{\textrm{sk}_i})^{-1}$\\ 
    $m = M.x - \Delta \pmod{q}$\\ 
    \textbf{return} $m$
\end{algorithm}

\subsection{Summary of Key Notations}

\begin{table}[h!]
\centering
\caption{Key Notations Used in the FDKG Protocol}
\label{tab:key_fdkg_notations} 
\begin{tabular}{cl}
\hline
\textbf{Notation} & \textbf{Description} \\
\hline
$n$ & Total number of potential parties in the system $\Parties$. \\
$\Parties$ & Set of all potential parties $\{\Party{1}, \dots, \Party{n}\}$. \\
$\Party{i}$ & Identifier for party $i$. \\
$\mathbb{G}, G, q$ & Cyclic group, its generator, and prime order. \\
$\textrm{pk}_i, \textrm{sk}_i$ & PKE public/secret key pair for $\Party{i}$. \\
$\textrm{crs}$ & Common Reference String for NIZK proofs. \\
$\pi$ & Generic symbol for a NIZK proof (e.g., $\pi_{FDKG_i}$). \\
\hline
$\GuardianSetOf{i}$ & Guardian set chosen by participant $\Party{i}$. \\
$t$ & Threshold for share reconstruction within a guardian set. \\
$k$ & Size of each participant's chosen guardian set. \\
$\SetOfFDKG$ & Set of valid parties in the generation phase. \\
$\PartialSecretKey{i}, \PartialPublicKey{i}$ & Partial secret/public key of participant $\Party{i} \in \SetOfFDKG$. \\
& ($\PartialPublicKey{i} = G^{\PartialSecretKey{i}}$). \\
$\SecretKey, \PublicKey$ & Global secret/public key aggregated from partial keys. \\
& ($\SecretKey = \sum \PartialSecretKey{i}$, $\PublicKey = \prod \PartialPublicKey{i} = G^{\SecretKey}$). \\
$s_{i,j}$ & SSS share of $\PartialSecretKey{i}$ for guardian $\Party{j} \in \GuardianSetOf{i}$. \\
$\EncryptedSharePartialSecretKey{i}{j}$ & Ciphertext of $s_{i,j}$ encrypted under $\textrm{pk}_j$. \\
$\textrm{Enc}(\textrm{pk}, \cdot, \cdot)$ & PKE encryption algorithm. \\
$\textrm{Dec}(\textrm{sk}, \cdot)$ & PKE decryption algorithm. \\
\hline
$\Voters$ & Set of parties participating in the voting phase. \\
$\Ballot{i}, \Vote{i}, \BlindingFactor{i}$ & Voter $\Party{i}$'s encrypted ballot, encoded vote, \\
& and blinding factor. \\
$\Tallies$ & Set of parties participating in the online tally phase. \\
$\TotalA, \TotalB$ & Aggregated ElGamal components of all cast ballots. \\
$\PartialDecryptionFrom{i}$ & Partial decryption of $\TotalA$ by $\Party{i}$ using $\PartialSecretKey{i}$. \\
$\SharePartialDecryptionFromTo{j}{i}$ & Partial decryption of $\TotalA$ by guardian $\Party{i}$ using share $s_{j,i}$.\\
\hline
\end{tabular}
\end{table}

Our implementation instantiates IND-CPA PKE with an ElGamal variant over BabyJub (Edwards curve) and NIZKs with Groth16. The DLP hardness assumption is therefore the discrete-log assumption in the BabyJub subgroup; IND-CPA security reduces to the DDH assumption in that group. Groth16 soundness/zero-knowledge rely on pairing-friendly curve security (BN254 in our implementation) and a trusted SRS for the given circuits. The security reductions in Section~\ref{sec:security_analysis} use only these abstract properties; other IND-CPA PKEs or NIZKs (PLONK/Marlin/STARKs) could be substituted with corresponding trade-offs in proof size and proving time.

\section{Federated Distributed Key Generation (FDKG)}
\label{sec:fdkg}

Federated Distributed Key Generation (FDKG) generalizes DKG to dynamic participation: any subset of parties can generate a joint key and delegate reconstruction to self-chosen guardian sets.

\subsection{Protocol Setting and Goal}
We consider a set $\Parties = \{\Party{1}, \dots, \Party{n}\}$ of $n$ potential parties. The system operates over a cyclic group $\mathbb{G}$ of prime order $q$ with generator $G$. We assume a Public Key Infrastructure (PKI) where each party $\Party{i}$ possesses a long-term secret key $\textrm{sk}_i$ and a corresponding public key $\textrm{pk}_i$ for an IND-CPA secure public-key encryption scheme $(\textrm{Enc}, \textrm{Dec})$ (Section~\ref{subsec:pke}), used for secure share distribution. Communication occurs via a broadcast channel. The protocol also utilizes a Non-Interactive Zero-Knowledge (NIZK) proof system $(\textrm{Setup}, \textrm{Prove}, \textrm{Verify})$ (Section~\ref{subsec:nizk}) which is assumed to be complete, computationally sound, and zero-knowledge, operating with a common reference string $\textrm{crs}$.

The goal of FDKG is for any subset of participating parties $\SetOfFDKG \subseteq \Parties$ to jointly generate a global public key $\PublicKey \in \mathbb{G}$ such that the corresponding secret key $\SecretKey \in \mathbb{Z}_q$ (where $\PublicKey = G^{\SecretKey}$) is shared among them. Specifically, each participant $\Party{i} \in \SetOfFDKG$ generates a partial key pair $(\PartialSecretKey{i}, \PartialPublicKey{i})$ where $\PartialPublicKey{i} = G^{\PartialSecretKey{i}}$. The global key pair is formed by aggregation: $\SecretKey = \sum_{i \in \SetOfFDKG} \PartialSecretKey{i}$ and $\PublicKey = \prod_{i \in \SetOfFDKG} \PartialPublicKey{i}$. Each participant $\Party{i}$ uses a local $(t, k)$-Shamir Secret Sharing (SSS) scheme (Section~\ref{subsec:sss}) to distribute its partial secret $\PartialSecretKey{i}$ among a self-selected guardian set $\GuardianSetOf{i} \subset \Parties \setminus \{\Party{i}\}$ of size $k$.

\subsection{The FDKG Protocol ($\Pi_{FDKG}$)}
The protocol proceeds in two rounds: Generation and Reconstruction.

\subsubsection{Round 1: Generation phase}
\label{subsubsec:fdkg_round1}

Any party $\Party{i} \in \Parties$ may (optionally) participate:
\begin{enumerate}
    \item \textbf{Guardian Selection:} Choose $\GuardianSetOf{i} \subset \Parties \setminus \{\Party{i}\}$ with $|\GuardianSetOf{i}| = k$.
    \item \textbf{Partial Key Generation:} Sample partial secret $\PartialSecretKey{i} \sample \mathbb{Z}_q$. Compute $\PartialPublicKey{i} = G^{\PartialSecretKey{i}}$.
    \item \textbf{Share Generation:} Generate shares $(s_{i,1}, \dots, s_{i,n}) \leftarrow \textrm{Share}(\PartialSecretKey{i}, t, n)$.
    \item \textbf{Share Encryption:} Initialize empty lists for ciphertexts $\mathbb{C}_i$ and randomness $\mathbb{R}_i$. For each $\Party{j} \in \GuardianSetOf{i}$:
        \begin{itemize}
            \item Sample PKE randomness: $(k_{i,j}, r_{i,j}) \sample \mathbb{Z}_q \times \mathbb{Z}_q$.
            \item Encrypt: $\EncryptedSharePartialSecretKey{i}{j} \leftarrow \textrm{Enc}_{\textrm{pk}_j}(s_{i,j}, [k_{i,j}, r_{i,j}])$.
            \item Store: Add $\EncryptedSharePartialSecretKey{i}{j}$ to $\mathbb{C}_i$ and add $(k_{i,j}, r_{i,j})$ to $\mathbb{R}_i$.
        \end{itemize}
    \item \textbf{Proof Generation:} Let the public statement be $x_i = (\PartialPublicKey{i}, \{\textrm{pk}_j\}_{\Party{j} \in \GuardianSetOf{i}}, \mathbb{C}_i)$. Let the witness be $w_i = (\PartialSecretKey{i}, \{s_{i,j}\}_{\Party{j} \in \GuardianSetOf{i}}, \mathbb{R}_i)$. Generate $\pi_{FDKG_i} \leftarrow \textrm{Prove}(\textrm{crs}, x_i, w_i)$ (see Appendix~\ref{app:proof-fdkg}).
    \item \textbf{Broadcast:} Broadcast $(\Party{i}, \PartialPublicKey{i}, \GuardianSetOf{i}, \mathbb{C}_i, \pi_{FDKG_i})$.
\end{enumerate}
        
\textbf{Post-Round 1 Processing (by any observer):}
\begin{itemize}
    \item Initialize $\SetOfFDKG = \emptyset$.
    \item For each received tuple $(\Party{i}, E_i, G_i, \mathbb{C}_i, \pi_i)$:
         \begin{itemize}
             \item Construct statement $x_i = (E_i, \{\textrm{pk}_j\}_{\Party{j} \in G_i}, \mathbb{C}_i)$. 
             \item Verify proof $b \leftarrow \textrm{Verify}(\textrm{crs}, x_i, \pi_i)$.
             \item If $b = 1$, add $\Party{i}$ to $\SetOfFDKG$ and store $(\PartialPublicKey{i}, \GuardianSetOf{i}, \mathbb{C}_i) \leftarrow (E_i, G_i, \mathbb{C}_i)$.
         \end{itemize}
    \item Compute $\PublicKey = \prod_{\Party{i} \in \SetOfFDKG} \PartialPublicKey{i}$.
    \item Public state: $\SetOfFDKG$, $\PublicKey$, $\{\PartialPublicKey{i}, \GuardianSetOf{i}, \mathbb{C}_i\}_{\Party{i} \in \SetOfFDKG}$.
\end{itemize}

\subsubsection{Round 2: Reconstruction Phase}
\label{subsubsec:fdkg_round2}

\textbf{Online Reconstruction (by parties $\Party{i} \in \Tallies \subseteq \Parties$):}
\begin{enumerate}
     \item \textbf{Reveal Partial Secret:} If $\Party{i} \in \SetOfFDKG \cap \Tallies$:
         \begin{itemize}
             \item Generate $\pi_{PS_i} \leftarrow \textrm{Prove}(\textrm{crs}, \PartialPublicKey{i}, \PartialSecretKey{i})$ (see Appendix~\ref{app:proof-ps}).
             \item Broadcast $(\Party{i}, \text{secret}, \PartialSecretKey{i}, \pi_{PS_i})$.
         \end{itemize}
     \item \textbf{Reveal Received Shares:} For each $\EncryptedSharePartialSecretKey{j}{i}$ (where $\Party{j} \in \SetOfFDKG, \Party{i} \in \GuardianSetOf{j}$):
         \begin{itemize}
             \item Decrypt share: $s_{j,i} \leftarrow \textrm{Dec}(\textrm{sk}_i, \EncryptedSharePartialSecretKey{j}{i})$.
             \item Generate proof: $\pi_{SPS_{j,i}} \leftarrow \textrm{Prove}(\textrm{crs}, (\EncryptedSharePartialSecretKey{j}{i}, \textrm{pk}_i, s_{j,i}), \textrm{sk}_i)$ (see Appendix~\ref{app:proof-sps}).
             \item Broadcast $(\Party{i}, \text{share}, \Party{j}, s_{j,i}, \pi_{SPS_{j,i}})$.
         \end{itemize}
\end{enumerate}

\textbf{Offline Reconstruction (by any observer):}
\begin{enumerate}
    \item Initialize empty list $D = \emptyset$.
    \item For each participant $\Party{i} \in \SetOfFDKG$:
        \begin{itemize}
            \item \textbf{Direct Secrets Reconstruction:} Look for $(\Party{i}, \text{secret}, \PartialSecretKey{i}, \pi_{PS_i})$. If $1 = \textrm{Verify}(\textrm{crs}, \PartialPublicKey{i}, \pi_{PS_i})$, add $\PartialSecretKey{i}$ to $D$;
            \item \textbf{Share-Based Reconstruction} \begin{enumerate}
                \item Initialize empty list $S_i$. For each $(\Party{k}, \text{share}, \Party{i}, s_{i,j}, \pi_{SPS_{i,j}})$ where $\Party{k}=\Party{j}$ and $\EncryptedSharePartialSecretKey{i}{j}$ exists: Verify $b \leftarrow \textrm{Verify}(\textrm{crs}, (\EncryptedSharePartialSecretKey{i}{j}, \textrm{pk}_j, s_{i,j}), \pi_{SPS_{i,j}})$. If $b=1$, add $(\Party{j}, s_{i,j})$ to $S_i$.
                \item Let $I_i = \{ \Party{j} \mid (\Party{j}, \cdot) \in S_i \}$. If $|I_i| \ge t$: Select $I'_i \subseteq I_i$, $|I'_i|=t$. Compute $\PartialSecretKey{i} \leftarrow \textrm{Reconstruct}(I'_i, \{s_{i,j} \mid (\Party{j}, s_{i,j}) \in S_i \text{ and } \Party{j} \in I'_i\})$. If $G^{\PartialSecretKey{i}} = \PartialPublicKey{i}$, add $\PartialSecretKey{i}$ to $D$.
            \end{enumerate}
        \end{itemize}
    \item \textbf{Compute Global Key:} If $D$ contains a valid entry for every $\Party{i} \in \SetOfFDKG$, compute $\SecretKey = \sum_{\PartialSecretKey{i} \in D} \PartialSecretKey{i}$. Output $\SecretKey$. Else output failure.
\end{enumerate}

\subsection{Communication Complexity}

The communication complexity depends on the number of participants $|\SetOfFDKG|$ and the size of guardian sets $k$.

\begin{itemize}
    \item \textbf{Round 1 (Generation):} Each participant $\Party{i} \in \SetOfFDKG$ broadcasts its partial public key $\PartialPublicKey{i}$ (1 group element), $k$ encrypted shares $\{\EncryptedSharePartialSecretKey{i}{j}\}$ (size depends on PKE scheme, in our scheme $k \times (2 |\mathbb{G}| + |\mathbb{Z}_q|)$), and one NIZK proof $\pi_{FDKG_i}$. Assuming $|\SetOfFDKG| \approx n$, the complexity is roughly $O(n k \cdot |\EncryptedSharePartialSecretKey{i}{j}| + n \cdot |\pi_{FDKG}|)$. In our BabyJub/Groth16 instantiation, this leads to $O(nk)$ complexity dominated by encrypted shares.
    \item \textbf{Round 2 (Online Reconstruction):} Each online party $\Party{i} \in \Tallies$ might broadcast its partial secret $\PartialSecretKey{i}$ (1 scalar) and proof $\pi_{PS_i}$, plus potentially many decrypted shares $s_{j,i}$ (1 scalar each) and proofs $\pi_{SPS_{j,i}}$. Let $m_i$ be the number of shares party $\Party{i}$ reveals (as a guardian for others). The complexity is roughly $O(|\Tallies| \cdot (|\mathbb{Z}_q| + |\pi_{PS_i}| + m_{max} \cdot (|\mathbb{Z}_q| + |\pi_{SPS_{j,i}}|)))$, where $m_{max} = \max_i m_i$. In the worst case, where $|\Tallies| \approx n$ and $m_{max} \approx n$, this can approach $O(n^2)$.
\end{itemize}
FDKG offers $O(n \cdot k)$ complexity for distribution, which is similar to standard DKG if $k \approx n$. Reconstruction complexity varies but can reach $O(n^2)$ in worst-case guardian—when a single party is selected as a guardian by every other participant and must reveal all $n-1$ shares.

\subsection{FDKG Example}
Consider $n=10$ potential parties $\{\Party{1}, \dots, \Party{10}\}$, local threshold $t=2$, and guardian set size $k=3$. Suppose $\SetOfFDKG = \{\Party{1}, \Party{3}, \Party{5}, \Party{7}, \Party{9}\}$ participate correctly in Round 1.

\begin{figure}[H]
    \centering
    \includegraphics[width=.5\textwidth]{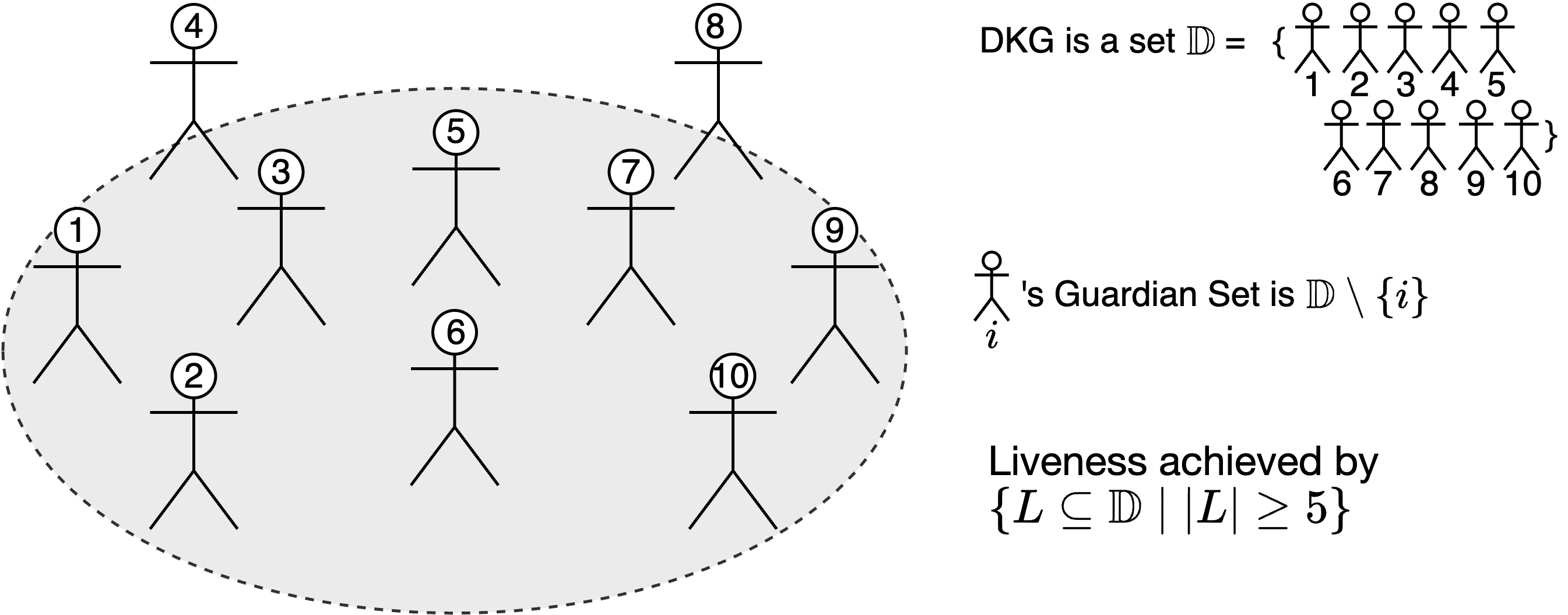} \\
    \vspace{0.5em} 
    \includegraphics[width=.5\textwidth]{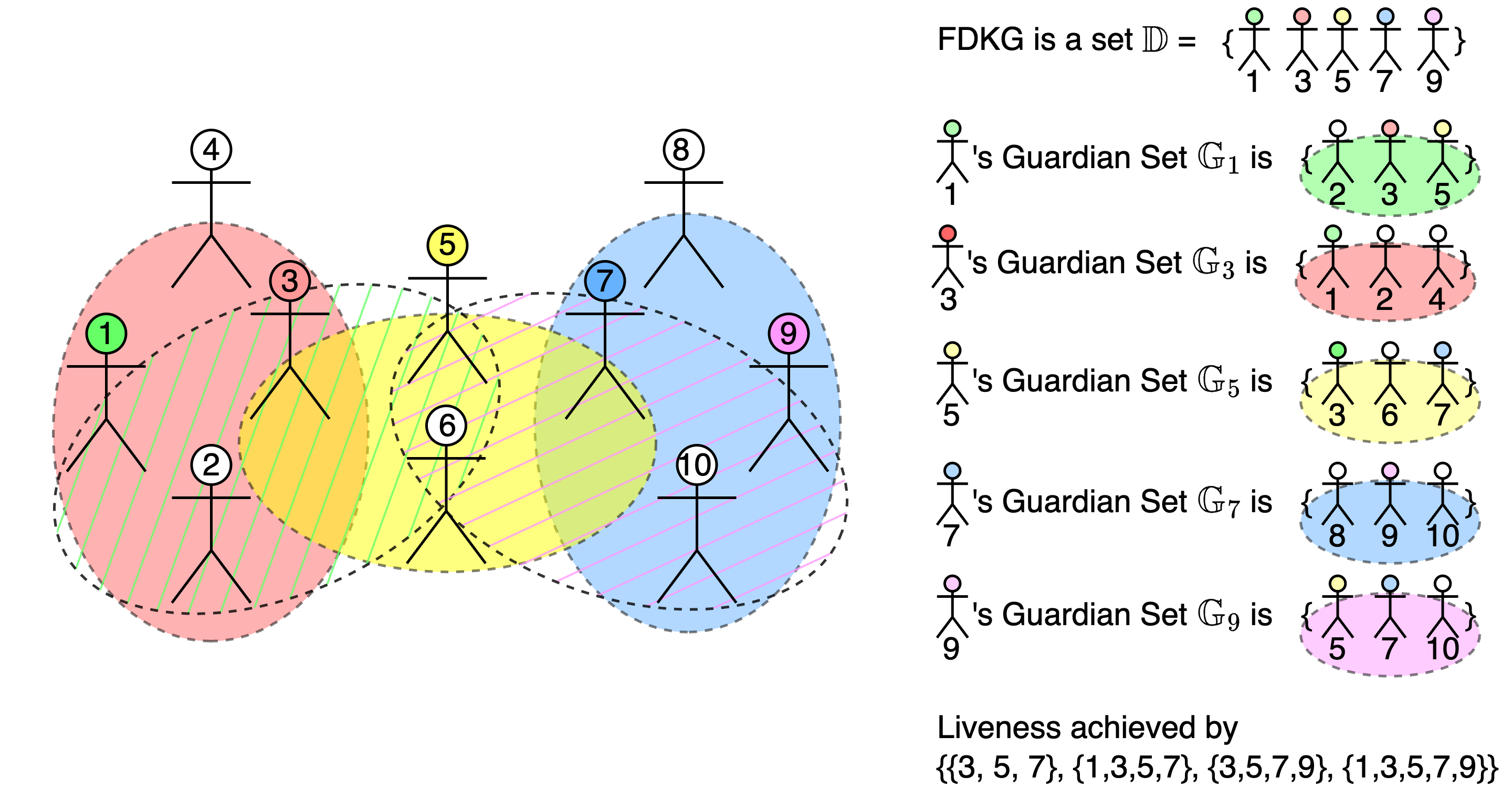}
    \caption{Visual comparison of $(k=10,t=5)$-DKG protocol (above) and $(n=10, k=3, t=2)$-FDKG (below)}
    \label{fig:FDKG}
\end{figure}

\begin{itemize}
    \item \textbf{Round 1 Online}
        \begin{itemize}
            \item $\Party{1}$ chooses $\GuardianSetOf{1} = \{\Party{2}, \Party{3}, \Party{5}\}$.
            \item Generates $(\PartialSecretKey{1}, \PartialPublicKey{1})$.
            \item Shares $\PartialSecretKey{1}$ into $s_{1,2}, s_{1,3}, s_{1,5}$.
            \item Encrypts shares to get $\mathbb{C}_1 = \{\EncryptedSharePartialSecretKey{1}{2}, \EncryptedSharePartialSecretKey{1}{3}, \EncryptedSharePartialSecretKey{1}{5}\}$.
            \item Generates $\pi_{FDKG_1}$ and broadcasts. Others in $\SetOfFDKG$ do similarly.
        \end{itemize}

    \item \textbf{Round 1 Offline}
        \begin{itemize}
            \item Once broadcasts are received, everyone computes $\PublicKey = \prod_{\Party{i} \in \SetOfFDKG} \PartialPublicKey{i}$.
        \end{itemize}

    \item \textbf{Round 2 Online}
        \begin{itemize}
            \item Suppose $\Tallies = \{\Party{3}, \Party{5}, \Party{7}\}$.
            \item $\Party{3}, \Party{5}, \Party{7}$ each broadcast their $(\text{secret}, \PartialSecretKey{\cdot}, \pi_{PS_{\cdot}})$.
            \item As guardians, they also decrypt and broadcast the necessary shares for offline parties:
                \begin{itemize}
                    \item $\Party{3}$ and $\Party{5}$ (guardians for $\Party{1}$) broadcast $(\text{share}, \Party{1}, s_{1,3}, \pi_{SPS_{1,3}})$ and $(\text{share}, \Party{1}, s_{1,5}, \pi_{SPS_{1,5}})$ respectively.
                    \item $\Party{5}$ and $\Party{7}$ (guardians for $\Party{9}$) broadcast $(\text{share}, \Party{9}, s_{9,5}, \pi_{SPS_{9,5}})$ and $(\text{share}, \Party{9}, s_{9,7}, \pi_{SPS_{9,7}})$ respectively.
                \end{itemize}
        \end{itemize}

    \item \textbf{Round 2 Offline}
        \begin{itemize}
            \item Collect all valid broadcasts: direct secrets from $\Party{3}, \Party{5}, \Party{7}$ and guardian shares for $\Party{1}$ and $\Party{9}$.
            \item Reconstruct offline parties using $t=2$ shares:
                \begin{itemize}
                    \item For $\Party{1}$: use $\{(\Party{3}, s_{1,3}), (\Party{5}, s_{1,5})\}$ to compute $\PartialSecretKey{1}' \leftarrow \textrm{Reconstruct}(\{\Party{3}, \Party{5}\}, \{s_{1,3}, s_{1,5}\})$ and verify $G^{\PartialSecretKey{1}'} = \PartialPublicKey{1}$.
                    \item For $\Party{9}$: use $\{(\Party{5}, s_{9,5}), (\Party{7}, s_{9,7})\}$ to compute $\PartialSecretKey{9}' \leftarrow \textrm{Reconstruct}(\{\Party{5}, \Party{7}\}, \{s_{9,5}, s_{9,7}\})$ and verify $G^{\PartialSecretKey{9}'} = \PartialPublicKey{9}$.
                \end{itemize}
            \item With $\PartialSecretKey{3}, \PartialSecretKey{5}, \PartialSecretKey{7}$ revealed directly and $\PartialSecretKey{1}', \PartialSecretKey{9}'$ reconstructed, compute $\SecretKey = \sum_{\Party{i} \in \SetOfFDKG} \PartialSecretKey{i}$.
        \end{itemize}
\end{itemize}

\section{Security Analysis}
\label{sec:security_analysis}

This section analyzes the security of $\Pi_{FDKG}$ in two phases: \emph{Generation} (Round~1) and \emph{Reconstruction} (Round~2).
We adopt the standard real/ideal paradigm~\cite{canettiUniversallyComposableSecurity2001, katzIntroductionModernCryptography2020}.
A detailed ideal functionality $\mathcal{F}_{FDKG}$ appears in Appendix~\ref{app:ideal_functionality}.

\subsection{Security Model and Assumptions}
\label{subsec:security_model}

\begin{enumerate}
    \item \textbf{Parties and communication.} There are $n$ potential parties $\Parties=\{\Party{1},\ldots,\Party{n}\}$. Communication occurs over an authenticated public broadcast channel in \emph{synchronous} rounds.
    \item \textbf{Adversary.} Unless stated otherwise, the adversary is \emph{static} probabilistic polynomial time (PPT) and corrupts a fixed subset $\Corrupted \subset \Parties$ before the protocol starts; corrupted parties are fully Byzantine. Let $\mathcal{H}=\Parties \setminus \Corrupted$ denote the honest set. Honest parties may be unavailable in Round~2 (crash/absence), which affects liveness. Section~\ref{subsec:adaptive_integrated} discusses adaptive adversaries and targeted attacks.
    \item \textbf{Cryptographic assumptions.}
    
    \begin{itemize}
        \item \textbf{PKE security.} The channel for encrypted shares $(\textrm{Enc}, \textrm{Dec})$ (Section~\ref{subsec:encrypted_channels}) is assumed \emph{IND-CPA} secure. In our implementation this is an ElGamal variant over the BabyJub subgroup. Security reduces to the \emph{DDH} assumption (and ultimately DLP hardness) in this group; ciphertext malleability is acceptable because correctness is enforced by NIZK soundness. Any alternative IND-CPA PKE could be substituted.

        \item \textbf{NIZK security.} The proof system $(\textrm{Setup}, \textrm{Prove}, \textrm{Verify})$ (Section~\ref{subsec:nizk}) is assumed computationally sound and zero-knowledge in the CRS model. We instantiate Groth16 over the \emph{BN254} pairing curve (also known as \emph{bn128}, the default in \texttt{circom}/\texttt{snarkjs}); this uses a circuit-specific structured reference string (SRS) generated in a trusted setup. Soundness relies on standard pairing-group assumptions on BN254; zero-knowledge follows from the Groth16 simulator. Other systems (e.g., PLONK/Marlin with a universal updatable SRS~\cite{gabizonPLONKPermutationsLagrangebases2019,chiesaMarlinPreprocessingZkSNARKs2019}, or STARKs with transparent setup~\cite{ben-sassonScalableTransparentPostquantum2018}) are compatible at the level of assumptions used here, trading proof size/proving time for setup properties and security level.

        \item \textbf{Discrete logarithm hardness.} We assume DLP hardness in (i) the BabyJub subgroup used for ElGamal and (ii) the pairing groups induced by the chosen SNARK curve, as applicable.
    \end{itemize}
\end{enumerate}

\subsection{Security of the Generation Phase (Round 1)}
\label{subsec:gen_security}

We prove Generation-phase security in the real/ideal paradigm using the ideal functionality $\mathcal{F}_{FDKG}$ (Appendix~\ref{app:ideal_functionality}).
Informally, the simulator replaces honest encrypted shares by encryptions of $0$ and uses the NIZK simulator to generate proofs; IND-CPA and zero-knowledge imply indistinguishability, while NIZK soundness enforces well-formedness. The following theorem summarizes the resulting properties.

\begin{theorem}[Generation: Correctness, Privacy, Robustness]
\label{thm:security_fdkg_generation_formal}
Assume the PKE is IND-CPA and the NIZK is zero-knowledge and computationally sound.
For the set of valid participants $\SetOfFDKG$ determined by verification:
\begin{enumerate}
\item \textbf{Correctness.}
For each $i\in\SetOfFDKG$, $\PartialPublicKey{i}=G^{\PartialSecretKey{i}}$ and
\[
\PublicKey \;=\; \prod_{i\in\SetOfFDKG}\PartialPublicKey{i}
\;=\; G^{\sum_{i\in\SetOfFDKG}\PartialSecretKey{i}}.
\]
Moreover, every published $\EncryptedSharePartialSecretKey{i}{j}$ encrypts $f_i(j)$
for a degree-$(t\!-\!1)$ polynomial $f_i$ with $f_i(0)=\PartialSecretKey{i}$.

\item \textbf{Privacy (during Generation).}
The adversary’s view leaks no information about
$\{\PartialSecretKey{i}\}_{i\in \SetOfFDKG\cap(\Parties\setminus\Corrupted)}$
beyond $(\PublicKey,\{\PartialPublicKey{i}\}_{i\in\SetOfFDKG})$ and the shares explicitly addressed to corrupted recipients
$\{\SharePartialSecretKey{i}{j}\mid j\in\GuardianSetOf{i}\cap\Corrupted\}$.

\item \textbf{Robustness (completion).}
Malformed broadcasts (invalid proofs) are rejected; corrupted parties cannot force
incorrect acceptance. Honest parties with valid broadcasts are included in
$\SetOfFDKG$ and obtain their outputs as per $\mathcal{F}_{FDKG}$.
\end{enumerate}
\end{theorem}

\begin{proof}[Proof sketch]
Simulate honest parties by encrypting $0$ and using simulated NIZKs; IND-CPA and zero-knowledge yield indistinguishability. NIZK soundness binds each dealer’s ciphertexts to a single polynomial and $\PartialPublicKey{i}$, ensuring well-formedness. Verification excludes malformed data, implying robustness.
\end{proof}

\subsection{Security of the Reconstruction Phase (Round 2)}
\label{subsec:recon_security}

This subsection assumes the Generation phase completed successfully (Theorem~\ref{thm:security_fdkg_generation_formal}) and that the Round~2 NIZKs ($\pi_{PS_i}$, $\pi_{SPS_{i,j}}$) are computationally sound. We analyze \emph{Correctness}, \emph{Privacy (reconstruction)}, and \emph{Liveness}.

\paragraph{Notation for Reconstruction}

Define the family of reconstruction-capable sets
\begin{equation}
\label{eq:R}
R \;:=\; \Bigl\{\, S \subseteq \Parties \ \Big|\ \forall i\in\SetOfFDKG:\ i\in S \ \ \lor\ \ |S\cap \GuardianSetOf{i}|\ge t \,\Bigr\}.
\end{equation}

Intuitively, $S\in R$ iff, for every participant $i$, $S$ contains either $i$ or at least $t$ of $i$’s guardians. Round~2 privacy and liveness depend on the guardian topology $\{\GuardianSetOf{i}\}$ via $R$ rather than on a single global corruption fraction, as in traditional uniform DKGs.

\begin{theorem}[Reconstruction: Correctness]
\label{thm:recon_correctness}
If for every $i\in\SetOfFDKG$ either (i) there is a valid broadcast $(\Party{i},\text{secret},\PartialSecretKey{i},\pi_{PS_i})$, or (ii) there exist $t$ distinct guardians $j\in\GuardianSetOf{i}$ with valid broadcasts $(\Party{j},\text{share},\Party{i},s_{i,j},\pi_{SPS_{i,j}})$, then any observer reconstructs the unique
\(
\SecretKey=\sum_{i\in\SetOfFDKG}\PartialSecretKey{i}.
\)
Moreover, checking $G^{\PartialSecretKey{i}}=\PartialPublicKey{i}$ prevents substitution.
\end{theorem}

\begin{proof}[Proof sketch]
By SSS, any $t$ correct shares interpolate $f_i(0)=\PartialSecretKey{i}$. Soundness of $\pi_{PS_i}$ and $\pi_{SPS_{i,j}}$ ensures that only values consistent with Round-1 commitments are accepted. Summing over $i$ yields $\SecretKey$.
\end{proof}

\begin{theorem}[Reconstruction: Privacy]
\label{thm:recon_privacy}
The adversary can reconstruct the global secret if and only if there exists a set $S\subseteq \Corrupted$ such that $S\in R$.
Equivalently, \emph{reconstruction privacy holds} iff $\Corrupted\notin R$.

\end{theorem}

\begin{proof}[Proof sketch]
If the adversary reconstructs, it must possess, for each $i$, either $i$’s partial secret (when $i\in\Corrupted$) or $t$ decryptable guardian shares (when $t$ guardians are corrupted). Collect those principals into $S$; then $S\subseteq\Corrupted$ and $S\in R$.

Conversely, if $S\subseteq\Corrupted$ and $S\in R$, then for every $i$ the adversary controls either $i$ or $t$ guardians. In the first case it knows $\PartialSecretKey{i}$; in the second it decrypts $t$ shares and reconstructs $\PartialSecretKey{i}$. Summing yields $\SecretKey$.
\end{proof}

\begin{theorem}[Reconstruction: Liveness]
\label{thm:recon_liveness}
The adversary cannot prevent reconstruction if and only if there exists a set $S\in R$ such that $S\subseteq U\setminus \Corrupted$.
Equivalently,
\[
\forall i\in\SetOfFDKG:\ \bigl(i\notin \Corrupted\bigr)\ \ \lor\ \ \bigl(|\Corrupted\cap \GuardianSetOf{i}| \le k-t\bigr).
\]
\end{theorem}

\begin{proof}[Proof sketch]
If such $S$ exists, then $S$ contains, for each $i$, either $i$ or $t$ honest guardians; reconstruction proceeds as in Theorem~\ref{thm:recon_correctness}. 

Conversely, if no such $S$ exists, then for some $i$ the adversary controls $i$ and at least $k-t+1$ of $\GuardianSetOf{i}$, preventing both direct revelation and a $t$-guardian reconstruction path; hence liveness fails.
\end{proof}

\paragraph*{Potential attacks and mitigations}
\emph{Share withholding / selective opening} in Round~2 is captured by Theorem~\ref{thm:recon_liveness}: liveness fails for $i$ only if the adversary both corrupts $i$ and at least $k{-}t{+}1$ of $\GuardianSetOf{i}$. \emph{Forgery or equivocation} is ruled out by soundness of $\pi_{PS_i}$ and $\pi_{SPS_{i,j}}$. \emph{Guardian concentration attacks} (attracting many delegations to a small set of guardians) increase redundancy under random churn but reduce privacy thresholds under adaptivity (Theorem~\ref{thm:recon_privacy}); deployers can mitigate by (i) capping per-guardian incoming delegations, (ii) diversifying recommendation lists, (iii) sampling a fraction of guardians via a verifiable random function (VRF), and (iv) rate-limiting or staking for high-degree guardians. Finally, \emph{Sybil guardians} are out of scope of our PKI model but can be addressed operationally via admission controls (identity or stake).

\paragraph*{On topology dependence}
Two extreme cases illustrate the role of overlap:
\begin{itemize}
\item \emph{Disjoint guardians.} If $\GuardianSetOf{i}\cap \GuardianSetOf{i'}=\varnothing$ for $i\neq i'$, then making a fixed $i$ unrecoverable requires at least $1+(k-t+1)$ targeted corruptions (participant $i$ and $k-t+1$ distinct guardians).
\item \emph{Full overlap.} If $\GuardianSetOf{i}=\GuardianSetOf{}$ for all $i$, then corrupting $k-t+1$ parties in $\GuardianSetOf{}$ makes every corrupted participant $i$ unrecoverable. Thus, the fraction of corruptions needed to break liveness or privacy depends on $\{\GuardianSetOf{i}\}$.
\end{itemize}
Next section empirically studies overlap policies (Random vs.\ Barabási–Albert) and their impact on success rates.

\section{Liveness Simulations}
\label{sec:liveness_simulations}

This section empirically evaluates \emph{liveness} under non-adversarial (random) churn. The goal is to estimate, for a configuration $(n,p,r,k,t)$ and a guardian-selection policy, the probability that the Round~2 reconstruction predicate induced by~\eqref{eq:R} is satisfied. These experiments complement the conditions in Theorems~\ref{thm:recon_liveness} and~\ref{thm:recon_privacy}: they do not constitute a cryptographic proof, but quantify the operational regimes where reconstruction is likely to succeed in the presence of availability failures.

\paragraph*{Evaluation predicate}
Let $\SetOfFDKG$ be the set of valid Round~1 participants and $\Tallies\subseteq\Parties$ be the Round~2 actors that are present. A trial is a \emph{success} iff
\begin{equation}
\label{eq:liveness_predicate}
\forall i \in \SetOfFDKG:\quad \bigl(i \in \Tallies\bigr)\ \ \lor\ \ \bigl(\bigl|\Tallies \cap \GuardianSetOf{i}\bigr| \ge t\bigr),
\end{equation}
i.e., iff $\Tallies\in R$ with $R$ as in~(\ref{eq:R}); see Theorem~\ref{thm:recon_liveness}.

\subsection{Experimental Design and Rationale}
\label{subsec:design_rationale}

\paragraph*{Topology policies (guardian selection)}
We study two canonical policies that bracket real deployments:

\begin{itemize}
  \item \textbf{Random (Erdős–Rényi, ER):} guardians are chosen uniformly; this models hub-free environments (uniform lists, randomized assignment).
  \item \textbf{Barabási–Albert (BA):} guardians are chosen with preferential attachment; this models environments with socially/operationally prominent ``hubs'' (recognizable, responsive, institutionally visible parties).
\end{itemize}

Both policies enforce $|\GuardianSetOf{i}| = k$ with local threshold $t\le k$. ER provides a neutral baseline without hubs; BA captures heavy-tailed guardian popularity, which concentrates redundancy around hubs and (under random churn) tends to increase the number of disjoint reconstruction paths.

\paragraph*{Parameters and sampling}
We explore
$n\in\{50,100,200,\dots,1000\}$,
$p\in\{0.1,0.2,\dots,1.0\}$,
$r\in\{0.1,0.2,\dots,1.0\}$,
$k\in\{1,3,\dots,n-1\}$,
and $t\in\{1,\dots,k\}$.
Round~1 participation draws $\SetOfFDKG$ with $|\SetOfFDKG|\approx p\,n$. Round~2 availability draws $\Tallies$ with expected size $\approx r\,|\Parties|$. Unless stated otherwise, we run 100 independent, identically distributed (i.i.d.) trials per configuration and report the \emph{success rate} (fraction of trials satisfying~\eqref{eq:liveness_predicate}).

\paragraph*{DKG vs FDKG}
Classical $(t,n)$-DKG is the special case of FDKG with full participation and maximal guardians:
\[
\text{DKG}(t,n) \equiv \text{FDKG}\bigl(p{=}1,\ t,\ k{=}n{-}1\bigr).
\]
Therefore, any comparison at identical parameters is tautological. In figures we include a \emph{reference line} labelled ``DKG'' defined as the FDKG curve at $(p{=}1,\ k{=}n{-}1)$; all other curves are FDKG with $(p\le1,\ k\le n{-}1)$. This makes explicit that differences arise from configuration, not from different protocols.

\begin{figure}[h]
    \centering
    \begin{minipage}{0.24\textwidth}
        \centering
        \includegraphics[width=\textwidth]{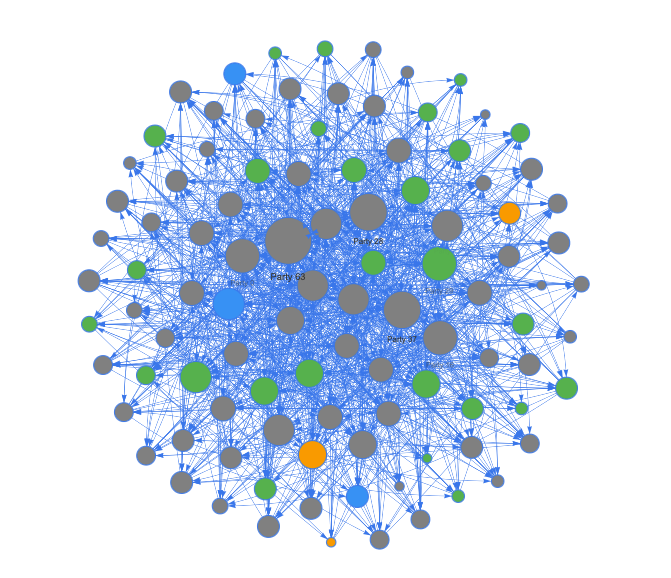}
    \end{minipage}
    \hfill
    \begin{minipage}{0.24\textwidth}
        \centering
        \includegraphics[width=\textwidth]{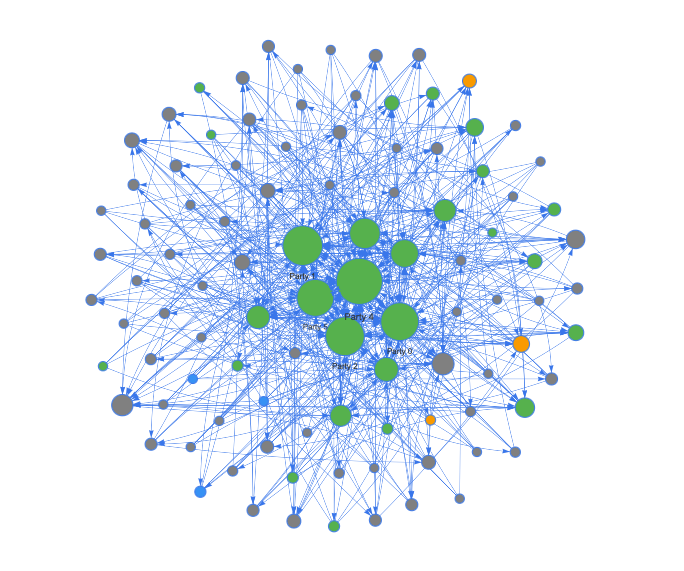}
    \end{minipage}
    \caption{ER (left) vs.\ BA (right) guardian-selection examples for $n=100, p=0.3, r=0.9, k=5, t=3$.
    Gray: absent; green: present in both rounds; blue: Round~1 only; orange: Round~2 only.}
    \label{fig:network_examples}
\end{figure}

\paragraph*{Implementation and artifacts}
We use the public simulator at \url{https://fdkg.stan.bar} (source: \url{https://github.com/stanbar/fdkg}). Figure~\ref{fig:network_examples} illustrates ER vs.\ BA for the same $(n,p,r,k,t)$.

\begin{figure}[htbp]
    \centering
    \includegraphics[width=0.5\textwidth]{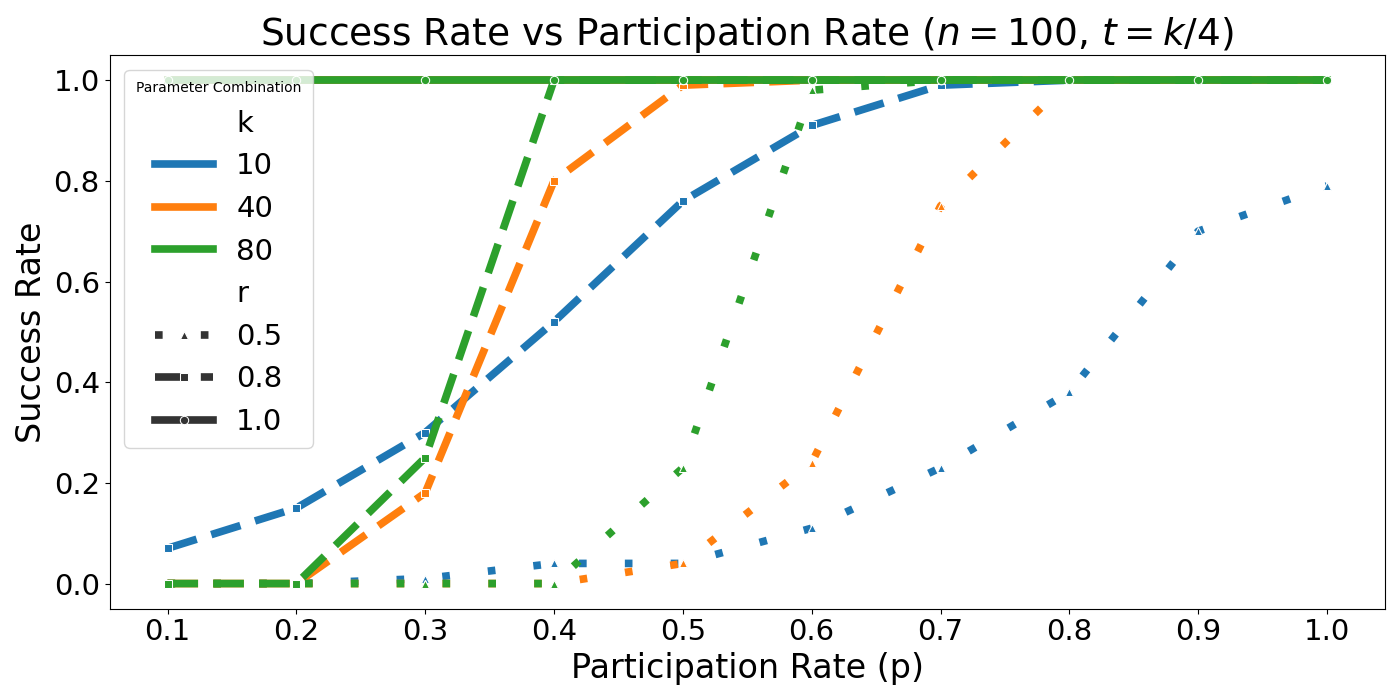}
    \includegraphics[width=0.5\textwidth]{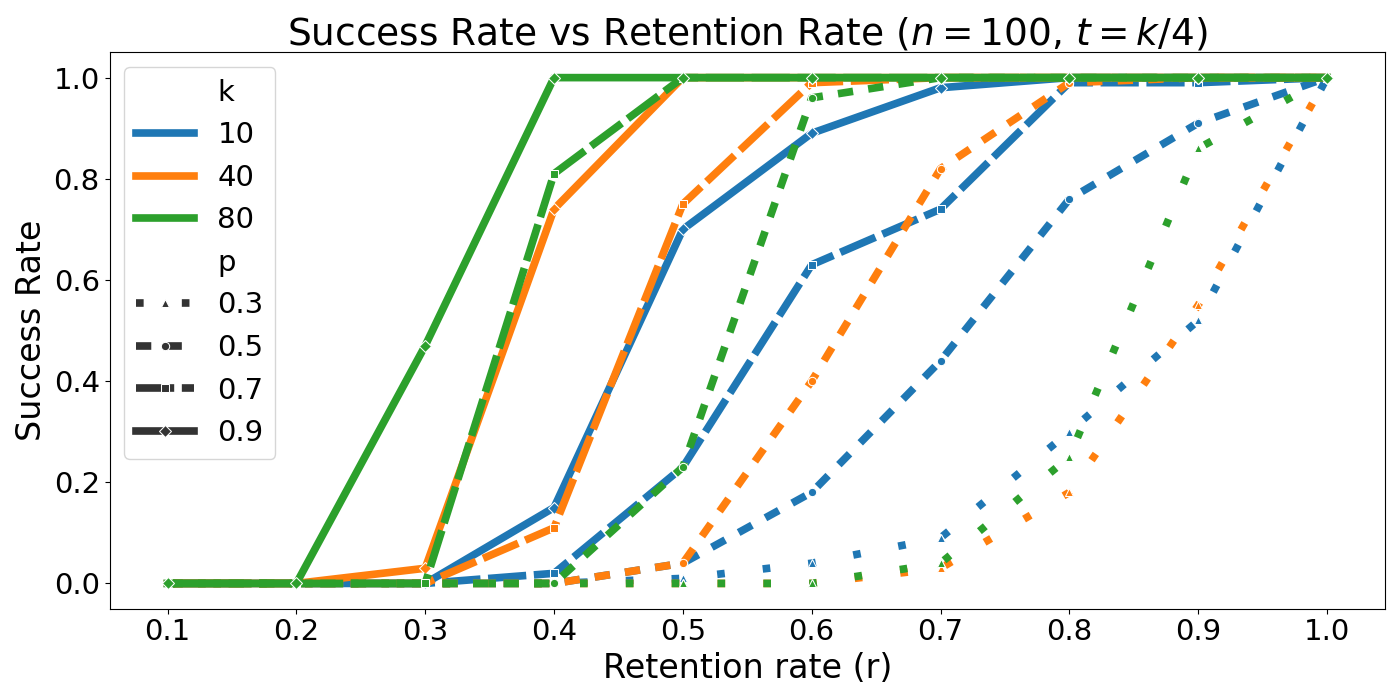}
    \caption{Success vs.\ participation $p$ (top) and retention $r$ (bottom) at $n=100$ with $t=k/4$. Line styles encode the other axis; colors encode $k$.}
    \label{fig:parameters_significance_participation}
\end{figure}

\begin{figure}[htbp]
    \centering
    \includegraphics[width=0.48\textwidth]{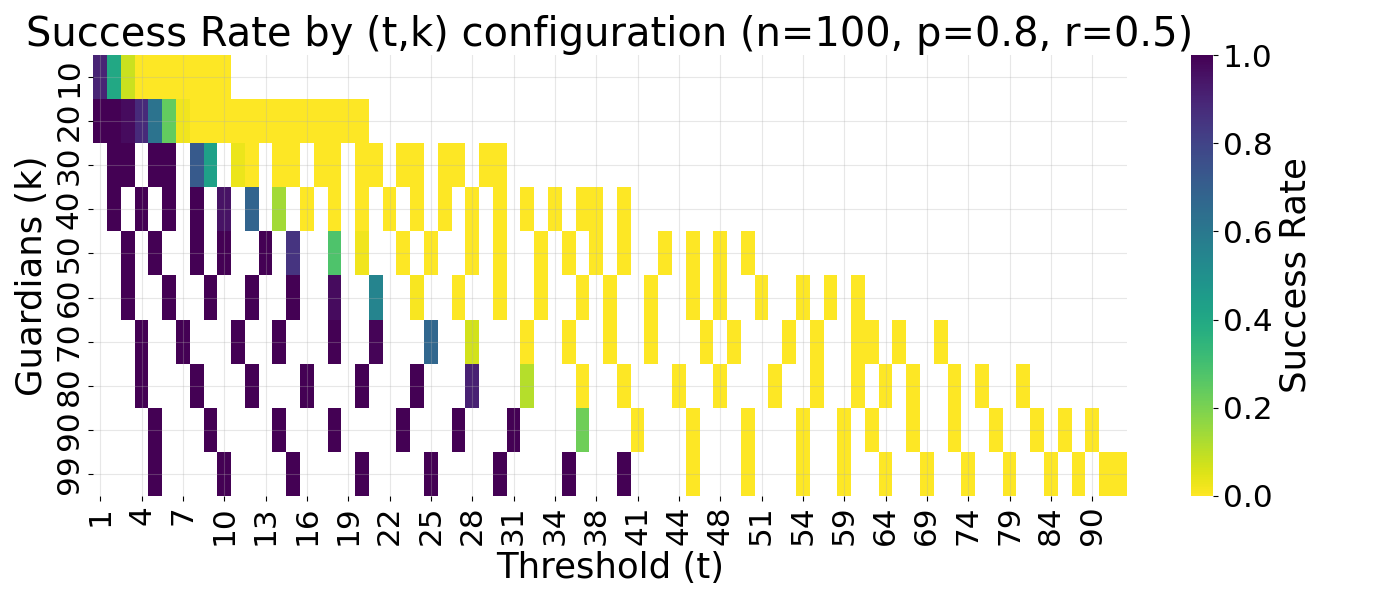}
    \includegraphics[width=0.48\textwidth]{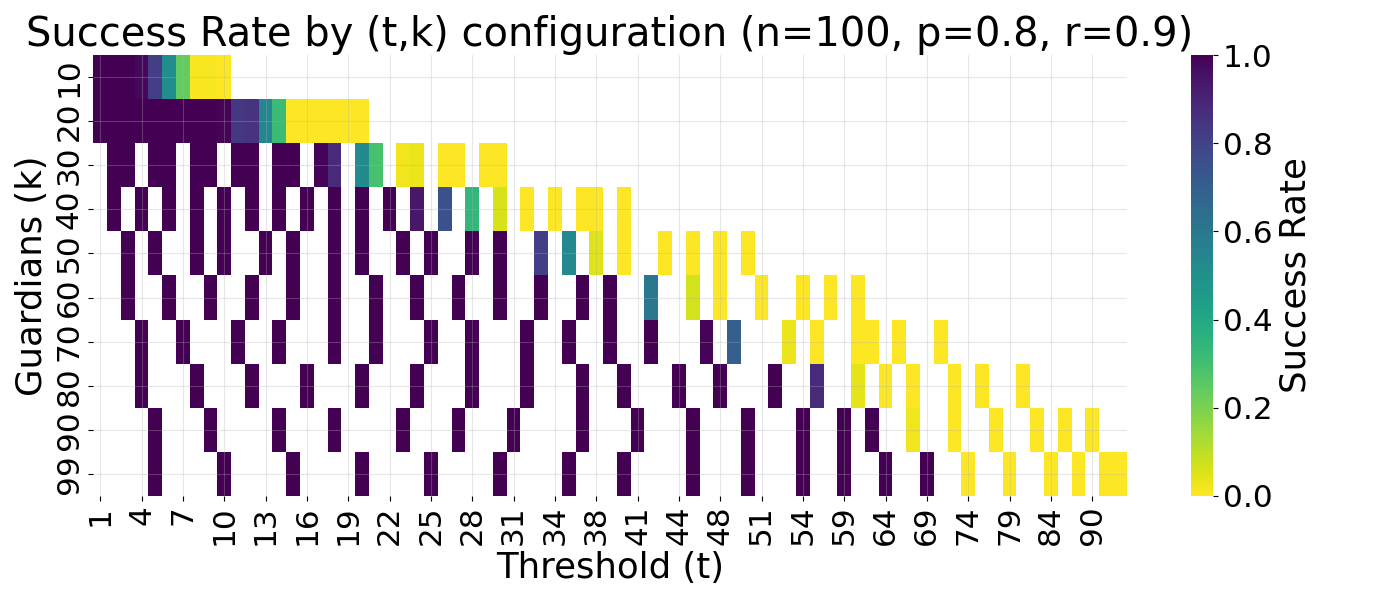}
    \caption{Success over $(k,t)$ for $n=100$, $p=0.8$, with $r=0.5$ (top) and $r=0.9$ (bottom). Darker indicates higher success. Higher $r$ expands the feasible region (larger $t$ at fixed $k$).}
    \label{fig:guardian_configs}
\end{figure}

\begin{figure}[htbp]
    \centering
    \includegraphics[width=0.5\textwidth]{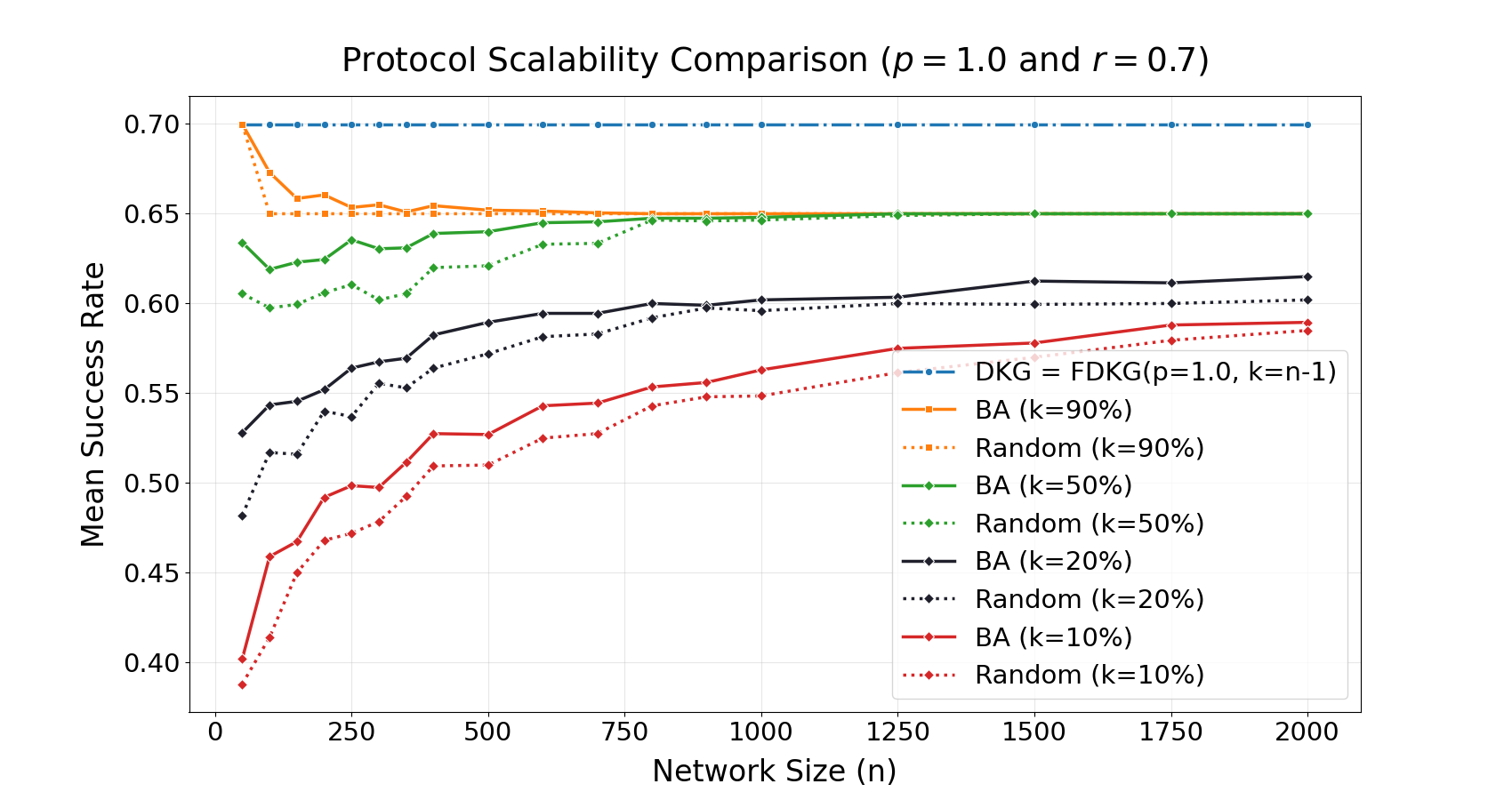}
    \caption{Scalability vs.\ guardian policy for $p=1.0,\ r=0.7$. The DKG line is the FDKG configuration $(p{=}1,\ k{=}n{-}1)$ (classical $(t,n)$-DKG). BA (solid) generally exceeds ER (dashed) under random churn; the gap narrows with larger $n$.}
    \label{fig:network_model}
\end{figure}

\subsection{Results and Interpretation}
\label{subsec:results}

\paragraph*{Retention dominates}
Retention $r$ is the primary driver of liveness. With $r\approx0.8$, high success is attainable even for modest $k$ at partial participation. For instance, for $n=100$, $p=0.7$, $k=10$ achieves high success. Increasng $k=40$ yields near-certainty even at $p=0.5$ (Figure~\ref{fig:parameters_significance_participation}).

With $r\approx0.5$, approaching certainty requires both higher $p$ and larger $k$ (e.g., for $n=100$, $p\gtrsim0.7$ and $k\gtrsim40$) (Figure~\ref{fig:parameters_significance_participation}).

\paragraph*{Threshold vs.\ redundancy}
For fixed $(n,p,r,k)$, success decreases as $t/k$ increases. Empirically, for $n=100$, $p=0.8$, $r=0.9$, liveness remains robust up to roughly $t \lesssim 0.7k$; with $r=0.5$ robustness holds up to roughly $t \lesssim 0.4k$ (Figure~\ref{fig:guardian_configs}).

\paragraph*{Topology under random churn}
BA generally outperforms ER for small/medium $n$ because preferential attachment concentrates redundancy around hubs, creating multiple disjoint reconstruction paths. As $n$ grows, the gap narrows (Figure~\ref{fig:network_model}).
This advantage pertains to \emph{random} availability failures and does not imply robustness to targeted hub failures.

\paragraph*{Scaling envelope}
Across $n$, the DKG line (FDKG at $p{=}1$, $k{=}n{-}1$) acts as an upper envelope at a given $r$. FDKG approaches this envelope as $k$ increases; for smaller $k$, it trades communication cost for liveness probability in a topology- and $r$-dependent manner.

\subsection{Adaptive Adversaries}
\label{subsec:adaptive_integrated}

The security analysis (Section~\ref{sec:security_analysis}) characterizes reconstruction privacy via $R$ (Theorem~\ref{thm:recon_privacy}) and liveness via the existence of an honest $S\in R$ (Theorem~\ref{thm:recon_liveness}). Under \emph{adaptive} corruption at Round~2, two objectives diverge:

\paragraph*{Targeting liveness}

To falsify~(\ref{eq:liveness_predicate}) for $i$, one must simultaneously prevent direct revelation (corrupt $i$) and reduce the remaining honest guardians below $t$ (thus corrupt $\ge k{-}t{+}1$ guardians).

\paragraph*{Targeting privacy}
Privacy fails iff the corrupted set contains a reconstruction-capable set for all participants' guardian requirements (Theorem~\ref{thm:recon_privacy}), i.e., there exists $S\subseteq \Corrupted$ with $S\in R$. 
The attacker needs to control enough guardians to reconstruct every participant's partial secret (unless that participant is directly corrupted). How many corruptions are needed depends mainly on how guardian sets overlap:
\begin{itemize}
  \item \emph{Full overlap} ($G_i=G$ for all $i$): corrupt any $t$ in $G$ to reconstruct \emph{all} partial secrets (recovers the classical DKG threshold).
  \item \emph{Disjoint guardians} ($G_i\cap G_j=\varnothing$): the adversary must, for each $i$, either corrupt $i$ or $t$ members of $G_i$, so the optimum scales with the number of participants unless many $i$ are directly corrupted.
\end{itemize}

Thus, diverse guardian choices improve availability under random churn (Section~\ref{subsec:results}), but privacy under targeted attacks depends on guardian-set overlaps. Operationally, implementations should provide UIs that encourage honest parties to choose diverse guardian sets to avoid over-concentration.

\section{Application: FDKG-Enhanced Voting Protocol}
\label{sec:voting_scheme}

In this section, we demonstrate an application of FDKG within an internet voting protocol. This protocol builds upon the foundational three-round structure common in threshold cryptosystems \cite{schoenmakersLectureNotesCryptographic2018}, incorporates multi-candidate vote encoding \cite{haoAnonymousVotingTworound2010}, and crucially utilizes FDKG (Section~\ref{sec:fdkg}) for decentralized and robust key management.

The core idea is to use FDKG to establish a threshold ElGamal public key $\PublicKey$ where the corresponding secret key $\SecretKey$ is distributed among participants and their guardians, mitigating single points of failure and handling dynamic participation.

The voting protocol consists of three primary rounds:
\begin{enumerate}
    \item \textbf{Key Generation (FDKG):} Participants establish the joint public encryption key $\PublicKey$ and distribute shares of their partial decryption keys.
    \item \textbf{Vote Casting:} Voters encrypt their choices under $\PublicKey$ and submit their ballots with proofs of validity.
    \item \textbf{Tallying:} Participants collaboratively decrypt the sum of encrypted votes to reveal the final tally.
\end{enumerate}

\subsection{Round 1: Federated Distributed Key Generation}

This round follows the $\Pi_{FDKG}$ protocol exactly as described in Section~\ref{subsubsec:fdkg_round1}.
\begin{itemize}
    \item \textbf{Participation:} Optional for any party $\Party{i} \in \Parties$.
    \item \textbf{Actions:} Each participating party $\Party{i}$ selects its guardian set $\GuardianSetOf{i}$, generates its partial key pair $(\PartialSecretKey{i}, \PartialPublicKey{i})$, computes and encrypts shares $\EncryptedSharePartialSecretKey{i}{j}$ for its guardians, generates a NIZK proof $\pi_{FDKG_i}$, and broadcasts its contribution.
    \item \textbf{Outcome:} Observers verify the proofs $\pi_{FDKG_i}$ to determine the set of valid participants $\SetOfFDKG$. The global public encryption key for the election is computed as $\PublicKey = \prod_{\Party{i} \in \SetOfFDKG} \PartialPublicKey{i}$. The public state includes $\SetOfFDKG$, $\PublicKey$, and the broadcasted tuples $\{\PartialPublicKey{i}, \GuardianSetOf{i}, \{\EncryptedSharePartialSecretKey{i}{j}\}, \pi_{FDKG_i}\}_{\Party{i} \in \SetOfFDKG}$.
\end{itemize}

\subsection{Round 2: Vote Casting}

Let $\Voters \subseteq \Parties$ be the set of eligible voters participating in this round. For each voter $\Party{i} \in \Voters$:
\begin{enumerate}
     \item \textbf{Vote Encoding:} Encode the chosen candidate into a scalar value $\Vote{i} \in \mathbb{Z}_q$. We use the power-of-2 encoding from \cite{baudronPracticalMulticandidateElection2001}:
     \[ \Vote{i}\ =\ \begin{cases} G^{2^0} & \text{for candidate 1} \\ G^{2^m} & \text{for candidate 2} \\ \vdots & \vdots \\ G^{2^{(c-1)m}} & \text{for candidate } c \end{cases} \]
     where $m$ is the smallest integer s.t. $2^m > n$, and $c$ is the number of candidates. The vote is represented as a point on the curve.
    \item \textbf{Vote Encryption (ElGamal):} Generate a random blinding factor $\BlindingFactor{i} \sample \mathbb{Z}_q$. Compute the encrypted ballot $\Ballot{i} = (\BallotA{i}, \BallotB{i})$ where:
    \begin{align*}
        \BallotA{i} &= G^{\BlindingFactor{i}} \\
        \BallotB{i} &= \PublicKey^{\BlindingFactor{i}} \cdot \Vote{i} \quad \text{(using group operation in } \G)
    \end{align*}
    \item \textbf{Proof Generation:} Generate a NIZK proof $\pi_{Ballot_i}$ demonstrating that $\Ballot{i}$ is a valid ElGamal encryption of a vote $\Vote{i}$ from the allowed set $\{G^{2^0}, G^{2^m}, \dots, G^{2^{(c-1)m}}\}$ under public key $\PublicKey$, without revealing $\Vote{i}$ or $\BlindingFactor{i}$. (See Appendix~\ref{app:proof-ballot}).
    \item \textbf{Broadcast:} Broadcast the ballot and proof $(\Ballot{i}, \pi_{Ballot_i})$.
\end{enumerate}

\paragraph*{State after Voting}
The broadcast channel is appended with the set of valid ballots and proofs:
\begin{itemize}
    \item $\{(\Ballot{i}, \pi_{Ballot_i}) \mid \Party{i} \in \Voters \text{ and } \pi_{Ballot_i} \text{ is valid}\}$. Let  $\Voters' \subseteq \Voters$ be the set of voters whose ballots verified.
\end{itemize}

\subsection{Round 3: Tallying}
This round involves threshold decryption of the combined ballots. It follows the structure of the FDKG Reconstruction Phase  (Section~\ref{subsubsec:fdkg_round2}).

\subsubsection{Online Tally}
A subset of parties $\Tallies \subseteq \Parties$ participate. For each party $\Party{i} \in \Tallies$:

\begin{enumerate}
    \item \textbf{Compute Homomorphic Sum:} Aggregate the first components of all valid ballots: $\TotalA = \prod_{\Party{j} \in \Voters'} \BallotA{j}$.
    
    \item \textbf{Provide Partial Decryption (if applicable):} If $\Party{i} \in \SetOfFDKG$:
        \begin{itemize}
            \item Compute the partial decryption: $\PartialDecryptionFrom{i} = (\TotalA)^{\PartialSecretKey{i}}$.
            \item Generate a NIZK proof $\pi_{PD_i}$ proving that $\PartialDecryptionFrom{i}$ was computed correctly using the secret $\PartialSecretKey{i}$ corresponding to the public $\PartialPublicKey{i}$. (See Appendix~\ref{app:proof-pd}).
            \item Broadcast $(\Party{i}, \text{pdecrypt}, \PartialDecryptionFrom{i}, \pi_{PD_i})$.
        \end{itemize}
        
    \item \textbf{Provide Decrypted Shares:} For each encrypted share $\EncryptedSharePartialSecretKey{j}{i}$ received in Round 1 (where $\Party{j} \in \SetOfFDKG$ and $\Party{i} \in \GuardianSetOf{j}$):
        \begin{itemize}
            \item Decrypt the share: $s_{j,i} = \textrm{Dec}_{\PartySecretKey{i}}(\EncryptedSharePartialSecretKey{j}{i})$.
            \item Compute the share's contribution to decryption: $\SharePartialDecryptionFromTo{j}{i} = (\TotalA)^{s_{j,i}}$.
            \item Generate a NIZK proof $\pi_{PDS_{j,i}}$ proving correct decryption of $\EncryptedSharePartialSecretKey{j}{i}$ to $s_{j,i}$ and correct exponentiation of $\TotalA$ by $s_{j,i}$. (See Appendix~\ref{app:proof-pds}).
            \item Broadcast $(\Party{i}, \text{pdecryptshare}, \Party{j}, \SharePartialDecryptionFromTo{j}{i}, \pi_{PDS_{j,i}})$.
        \end{itemize}
\end{enumerate}

\paragraph*{State after Online Tally}
The broadcast channel contains potentially revealed partial decryptions $(\PartialDecryptionFrom{i}, \pi_{PD_i})$ and shares of partial decryptions $(\SharePartialDecryptionFromTo{j}{i}, \pi_{PDS_{j,i}})$ from parties in $\Tallies$.

\subsubsection{Offline Tally}
Performable by any observer using publicly available data.
\begin{enumerate}
    \item \textbf{Aggregate Ballot Components:} Compute the aggregated components of valid ballots:
        \begin{align*}
         \TotalA &= \prod_{\Party{k} \in \Voters'} \BallotA{k} \\
         \TotalB &= \prod_{\Party{k} \in \Voters'} \BallotB{k}
        \end{align*}
    \item \textbf{Reconstruct Partial Decryptions:} For each $\Party{i} \in \SetOfFDKG$:
        \begin{itemize}
            \item Try to find a valid broadcast $(\Party{i}, \text{pdecrypt}, \PartialDecryptionFrom{i}, \pi_{PD_i})$. If found and $\pi_{PD_i}$ verifies, use $\PartialDecryptionFrom{i} = \PartialDecryptionFrom{i}$.

            \item Otherwise, collect all valid broadcasted shares $(\Party{k}, \text{pdecryptshare}, \Party{i}, \SharePartialDecryptionFromTo{i}{j}, \pi_{PDS_{i,j}})$ where $\Party{k}=\Party{j}$ and $\pi_{PDS_{i,j}}$ verifies. Let $I_i$ be the set of indices $j$ for which valid shares were found.

            \item If $|I_i| \ge t$: Select $I'_i \subseteq I_i$, $|I'_i|=t$. Use Lagrange interpolation in the exponent to compute $\PartialDecryptionFrom{i} = \prod_{\Party{j} \in I'_i} (\SharePartialDecryptionFromTo{i}{j})^{\lambda_j}$. Verify this reconstructed value against $\PartialPublicKey{i}$.
            \item If neither direct secret nor enough shares are available/valid, the tally fails.
        \end{itemize}
    \item \textbf{Combine Partial Decryptions:} Compute the combined decryption factor $Z$:
        \[ Z = \prod_{\Party{i} \in \SetOfFDKG} \PartialDecryptionFrom{i} = (\TotalA)^{\sum \PartialSecretKey{i}} = (\TotalA)^{\SecretKey} \]
    \item \textbf{Decrypt Final Result:} Compute the product of the encoded votes:
        \[ M = \TotalB \cdot Z^{-1} = \left( \prod_{\Party{k} \in \Voters'} \PublicKey^{\BlindingFactor{k}} \cdot \Vote{k} \right) \cdot \left( \prod_{\Party{k} \in \Voters'} G^{\BlindingFactor{k}} \right)^{-\SecretKey} \]
        Since $\PublicKey = G^{\SecretKey}$, this simplifies to:
        \[ M = \prod_{\Party{k} \in \Voters'} \Vote{k} = G^{\sum x_k 2^{(k-1)m}} \]
        where $x_k$ is the number of votes for candidate $k$.
    \item \textbf{Extract Counts:} Solve the discrete logarithm problem $M = G^{\text{result}}$. Since `result` = $\sum x_k 2^{(k-1)m}$ and the total number of votes $|\Voters'|$ is known (bounding the $x_k$), the counts $x_k$ can be extracted efficiently, e.g., using Pollard's Rho or Baby-Step Giant-Step for the small resulting exponent, or simply by examining the binary representation if $m$ is chosen appropriately \cite{haoAnonymousVotingTworound2010}.
\end{enumerate}

\section{Performance Evaluation}
\label{sec:performance_evaluation}

This section presents a performance analysis of the FDKG-enhanced voting protocol described in Section~\ref{sec:voting_scheme}. We evaluated computational costs (NIZK proof generation times and offline tally time) and communication costs (total broadcast message sizes). All NIZK evaluations use Groth16 \cite{grothSizePairingBasedNoninteractive2016} as the NIZK proof system, assuming a proof size of 256 bytes (uncompressed). We briefly investigated using the PLONK \cite{gabizonPLONKPermutationsLagrangebases2019} proof system as an alternative potentially avoiding a trusted setup, but found it computationally infeasible for our circuits; for instance, generating a $\pi_{FDKG_i}$ proof with parameters $(t=3, k=10)$ took over 8 minutes with PLONK, and larger configurations exceeded available memory resources. Therefore, we proceeded with Groth16.

\paragraph{Computational and Communication Costs per Message Type}
Table~\ref{table:perf_summary} summarizes the core performance metrics for the different messages broadcast during the online phases of the protocol. It includes the time to generate the required NIZK proof and the total message size, which comprises the cryptographic payload plus the 256-byte Groth16 proof.

\begin{table*}
    \centering
    \caption{Performance Summary: NIZK Proving Time and Total Message Size per Broadcast Message Type.}
    \label{table:perf_summary} 
    \begin{tabular}{|l|c|c|c|c|c|c|}
    \hline
        \multirow{2}{*}{\textbf{\shortstack{Measurement}}} & \multicolumn{3}{c|}{\textbf{FDKG Generation}}  & \multirow{2}{*}{\textbf{\shortstack{Ballot \\ Broadcast}}} & \multirow{2}{*}{\textbf{\shortstack{Partial Dec. \\ Broadcast}}} & \multirow{2}{*}{\textbf{\shortstack{Part. Dec. Share \\ Broadcast}}} \\ 
        \cline{2-4}
        & \textbf{(t=3, k=10)} & \textbf{(t=10, k=30)} & \textbf{(t=30, k=100)} & & & \\ 
        \hline
        \textbf{NIZK Proving Time} & 3.358 s & 4.415 s & 14.786 s & 0.211 s & 0.619 s & 0.580 s/share \\ \hline
        \textbf{Base Payload Size} & 1664 B & 4864 B & 16064 B & 128 B & 64 B & 64 B/share \\
        \textbf{NIZK Proof Size} & 256 B & 256 B & 256 B & 256 B & 256 B & 256 B/share \\ \hline
        \textbf{Total Message Size (Bytes)} & \textbf{1920} & \textbf{5120} & \textbf{16320} & \textbf{384} & \textbf{320} & \textbf{320/share} \\
        \hline
    \end{tabular}
\end{table*}

The online computational cost is dominated by the generation of the $\pi_{FDKG_i}$ proof, especially as $t$ and $k$ increase. The communication cost for the FDKG Generation phase is primarily driven by $k$. Messages related to voting and tallying are significantly smaller individually, but the total communication cost of tallying depends heavily on how many secrets and shares need to be revealed publicly.

\paragraph{Offline Tally: Vote Extraction Cost}
\label{subsec:offline_tally_cost}
The final step of the Offline Tally phase (Section~\ref{sec:voting_scheme}, Offline Tally, Step 5) involves extracting the individual vote counts $\{x_k\}$ from the aggregated result $M = G^{\sum x_k 2^{(k-1)m}}$. This requires solving a Discrete Logarithm Problem (DLP) where the exponent has a specific known structure. Since the maximum value of the exponent is bounded by the total number of valid votes $|\Voters|$, standard algorithms like Baby-Step Giant-Step or Pollard's Rho can be used.

\begin{figure}[h] 
    \centering
    \includegraphics[width=.5\textwidth]{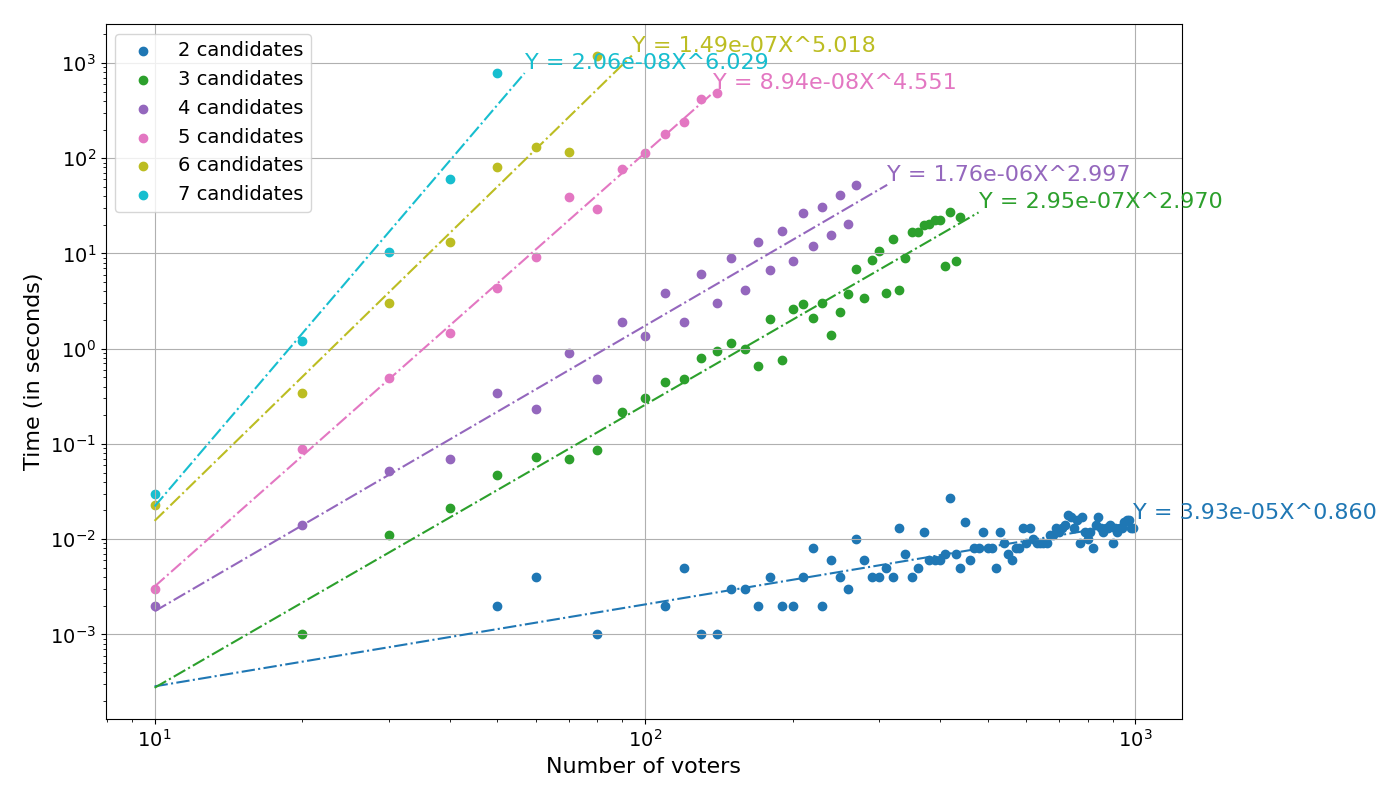} 
    \caption{Time required to solve DLP for vote extraction with respect to the number of candidates and number of voters (log-log scale).}
    \label{fig:dlog-search}
\end{figure}

We measured the time required for this vote extraction step for varying numbers of voters $|\Voters|$ and candidates $c$. The results are presented in Figure~\ref{fig:dlog-search}. The data shows that the time required grows approximately linearly with the number of voters (for a fixed number of candidates) but grows exponentially with the number of candidates $c$. For instance, with 2 candidates, extracting results for up to 1000 voters takes only milliseconds. However, with 5 candidates, extracting the result for 100 voters took over 100 seconds.

This indicates that while the FDKG and online tallying phases scale well, the vote extraction step using this specific power-of-2 encoding \cite{baudronPracticalMulticandidateElection2001} imposes a practical limit on the number of candidates that can be efficiently supported, especially when combined with a large number of voters.

\paragraph{Example Scenario Calculation}
\label{subsec:example_scenario} 

Let's estimate the total broadcast communication size for a specific scenario:
\begin{itemize}
    \item $n=100$ potential participants.
    \item $|\SetOfFDKG|=50$ valid participants in Round 1 (FDKG Generation).
    \item Each participant uses $k=40$ guardians.
    \item $|\Voters|=50$ valid ballots cast in Round 2.
    \item In Round 3 (Online Tally), assume $|\Tallies \cap \SetOfFDKG| = 40$ participants reveal their secrets directly.
    \item Assume for reconstruction, each of the 40 online talliers reveals, on average, 40 shares, this leads to approximately $40 \times 40 = 1600$ total shares being revealed (this is a simplification, the actual number depends on guardian set overlaps and the specific set of online talliers).
\end{itemize}

Using the "Total Message Size (Bytes)" figures from Table~\ref{table:perf_summary}:

\begin{enumerate}
    \item \textbf{FDKG Dist.:} $50 \times (64 + 40 \times 160 + 256) = 50 \times 6720 \text{ B} = 336000$ B (336.0 kB).
    \item \textbf{Voting:} $50 \times 384 \text{ B} = 19200$ B (19.2 kB).
    \item \textbf{Tallying (Online):}
        \begin{itemize}
            \item Secrets: $40 \times 320 \text{ B} = 12800$ B (12.8 kB).
            \item Shares: $1600 \times 320 \text{ B} = 512000$ B (512.0 kB).
        \end{itemize}
    \item \textbf{Total Broadcast:} $336000 + 19200 + 12800 + 512000 = 880000$ B (\textbf{880.0 kB}).
\end{enumerate}
The offline tally time for this scenario (50 voters) would be milliseconds for 2-4 candidates, but could increase to several minutes for 5 or more candidates, as indicated by Figure~\ref{fig:dlog-search}.

\paragraph*{Large-scale scenario ($n=5000$)}
Assume $|\SetOfFDKG|=2500$, $k=40$, $|\Voters|=2500$, $|\Tallies\cap\SetOfFDKG|=2000$, and $40$ shares revealed per tallier ($80{,}000$ shares total). Using Table~\ref{table:perf_summary} sizes:
\[
\begin{aligned}
\text{FDKG Dist.} & = 2500\times 6720 = 16{,}800{,}000~\text{B}~(16.8~\text{MB}),\\
\text{Voting} & = 2500\times 384 = \phantom{0}960{,}000~\text{B}~(0.96~\text{MB}),\\
\text{Tally (secrets)} & = 2000\times 320 = \phantom{0}640{,}000~\text{B}~(0.64~\text{MB}),\\
\text{Tally (shares)} & = 80{,}000\times 320 = 25{,}600{,}000~\text{B}~(25.6~\text{MB}),\\
\textbf{Total} & = 44{,}000{,}000~\text{B}~(\textbf{44.0~MB}).
\end{aligned}
\]
Communication is dominated by share revelations; for fixed $k$ and average shares per tallier, growth is near-linear in $|\SetOfFDKG|$.

\paragraph{Scalability and load balance}
FDKG uses one broadcast round in Generation and one in Reconstruction. End-to-end latency is dominated by (i) local proving time for $\pi_{FDKG_i}$ and (ii) network propagation per round; Reconstruction latency additionally depends on the number of shares each online guardian reveals and proves. Load is uneven by design: some guardians receive more delegations and therefore handle more decryption-share work. Let $c_g$ denote the number of parties that selected guardian $g$ (guardian “popularity”). A simple way to summarize skew is to report $\max_g c_g$, the median of $\{c_g\}$, and the ratio $\max_g c_g / \mathrm{median}(\{c_g\})$. In our experiments, this ratio is larger under BA than ER (reflecting hubs). Operators can bound peak load by capping per-guardian assignments, randomizing a fraction of guardian picks, or recommending diversified guardian sets.

The Generation broadcast cost scales as $\mathcal{O}(n\cdot k)$ elements (Section~\ref{sec:fdkg}); with fixed $k$ the total cost is linear in $n$. Reconstruction can reach $\mathcal{O}(n^2)$ in worst-case guardian topologies (some parties chosen by all other participants), but in practice remains near $\mathcal{O}(n\cdot k)$ when overlap is moderate. Hierarchical deployments (partitioning parties into domains, then aggregating domain-level partial keys) can reduce the maximum number of counterparties any single party must send to or serve during Generation/Reconstruction, while preserving heterogeneous trust.

\section{Deployments}
\label{sec:deployments}
\begin{figure}[h]
    \centering
    \includegraphics[width=0.5\textwidth]{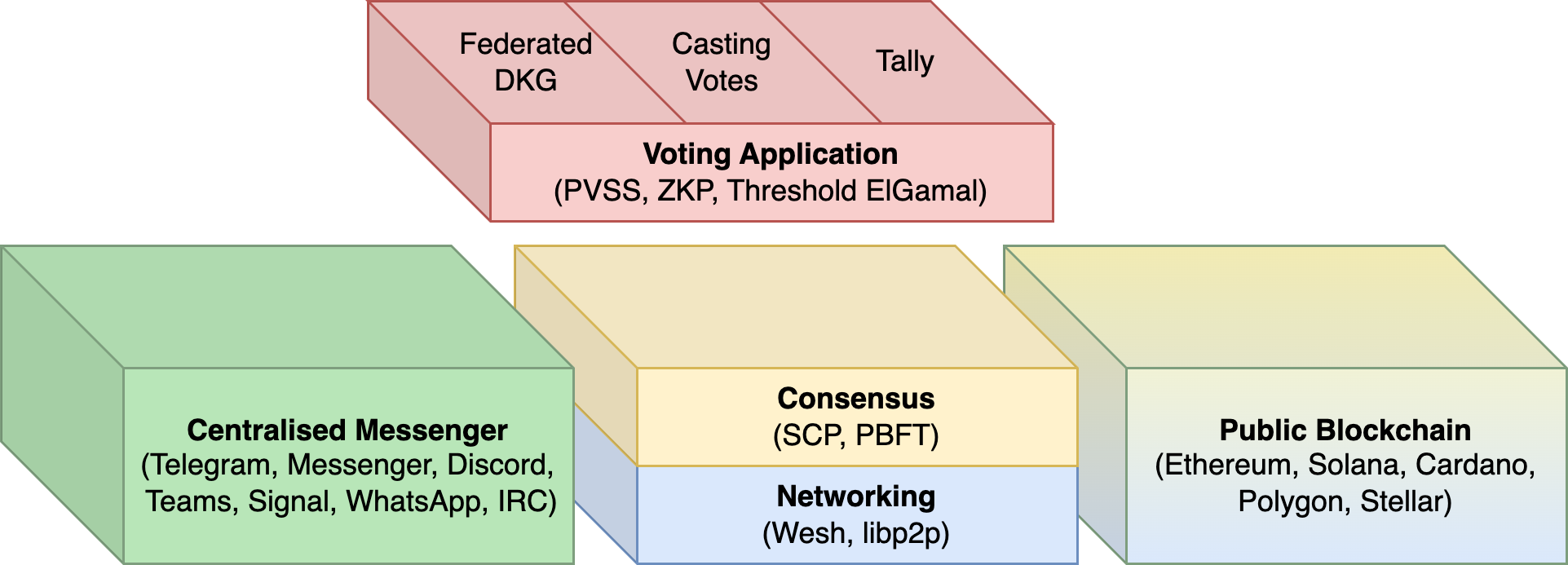}
    \caption{Three possible deployments of the protocol: Centralised messenger, ad-hoc peer-to-peer network, and public blockchain.}
    \label{fig:stack-bc}
\end{figure}

The FDKG protocol (and FDKG-based voting application), requiring only a shared communication space (message board) where participants can publicly post and read messages, is adaptable to various deployment environments. We consider three illustrative scenarios:

\begin{enumerate}
    \item \textbf{Public Blockchains:} The protocol can be implemented on public blockchains. For instance, interactions could be managed via smart contracts. Features like ERC-4337 (Account Abstraction)~\cite{ERC4337AccountAbstraction} could further simplify deployment by enabling centralized payment of transaction fees, potentially lowering adoption barriers for participants.
    \item \textbf{Peer-to-peer Networks:} The protocol is inherently suitable for peer-to-peer (P2P) networks. Technologies such as Wesh Network\footnote{Wesh Network, an asynchronous mesh network protocol by Berty Technologies, \url{https://wesh.network/}}, which provide decentralized communication infrastructure, can serve as a foundation for FDKG deployments.
    \item \textbf{Centralized Messengers:} For ease of deployment and accessibility, FDKG could be layered on top of existing centralized messenger platforms (e.g., Telegram, Signal, WhatsApp), where a group chat acts as the message board. However, it is important to note that the centralized nature of the messenger could introduce a trusted third party capable of censorship, potentially impacting protocol liveness or fairness if it selectively blocks messages.
\end{enumerate}
Figure~\ref{fig:stack-bc} visually depicts these potential deployment architectures.

\section{Limitations and Future Work}
\label{sec:limitations-and-future}
While FDKG offers enhanced flexibility and resilience to dynamic participation, our analysis and implementation highlight areas for further improvement and research.

\textbf{Cryptographic Overhead:} The protocol inherits computational overhead from its cryptographic building blocks, particularly NIZK proofs. Although Groth16 proofs, as used in our implementation, are efficient for mid-scale settings (e.g., tens to a few hundred participants), very large-scale or time-critical applications (like high-throughput elections) might necessitate faster proving systems (e.g., different SNARKs or STARKs) or techniques for proof aggregation to reduce verification load.

\textbf{Offline Tallying Scalability:} As discussed in Section~\ref{sec:performance_evaluation}, the offline tallying step in our FDKG-enhanced voting example, specifically the brute-force decoding of the final aggregated vote via DLP solution, exhibits exponential scaling with the number of candidates. Developing more efficient multi-candidate vote encoding schemes or advanced number-theoretic techniques for vote extraction remains an important open research challenge for such applications.

\textbf{Usability and guardian selection:}
Non-experts may find manual guardian selection burdensome. In practice, UIs can (i) offer curated recommendations that blend centrality (availability) with diversity (to avoid over-concentration), (ii) enforce soft caps on per-guardian assignments, (iii) set $(k,t)$ defaults from observed retention (e.g., $t\approx 0.3$–$0.5$ of $k$ for moderate $r$), and (iv) warn about privacy risks under highly overlapping guardian choices. These controls preserve FDKG’s heterogeneous trust while reducing user burden.

\textbf{Trusted Setup for NIZKs:} Our current NIZK instantiation (Groth16) requires a trusted setup for its Common Reference String (CRS). While this is a one-time cost per circuit, it represents a potential point of centralization or trust. Future work could explore integrating FDKG with NIZK systems that have transparent or universal setups (e.g., STARKs, or SNARKs like PLONK with a universal updatable SRS) to mitigate this dependency, although this may involve performance trade-offs.

\textbf{Resharing and forward secrecy:} FDKG, as presented, is a bootstrapping primitive. Enabling long-lived deployments with proactive defense against key accumulation requires a dynamic resharing protocol that (i) changes guardian sets $G_i$ and/or shares without re-running Generation, and (ii) provides forward secrecy against state compromise. Integrating proactive secret sharing (DPSS) with heterogeneous guardian sets while preserving public verifiability and succinct proofs is an open direction.

\textbf{Adaptive and targeted adversaries:}
So far the analysis assumes static corruption and random churn. Future work might study \emph{adaptive/rushing} behavior and \emph{targeted removals} against the guardian topology \(\{G_i\}\).
The goal is to quantify how privacy and liveness degrade when the attacker selects corruptions after seeing Round~1 transcripts, and to assess practical mitigations: mild commit--reveal or public randomness to limit key bias, plus topology-aware defenses.
Using the same reconstruction predicate \(R\), one could report topology-sensitive lower bounds on the effort needed to break privacy or liveness and identify when targeted strategies matter in practice.

\textbf{Topology and load control:}
Hotspots and concentration risks can be reduced by combining local diversity constraints (soft/hard caps per guardian, randomly sampled guardians, overlap-distance limits, reputation that discounts hubs) with a hierarchical FDKG where domains run local FDKG and a second tier aggregates domain keys.
Future work might measure how these knobs reshape \(\{G_i\}\): lowering \(\max_g c_g / \mathrm{median}(c_g)\), bounding worst-case reconstruction traffic, and increasing the minimal reconstruction-capable set for targeted breaks---while preserving heterogeneous trust and near \(O(nk)\) per tier.

\textbf{Broader Applications:} Although this paper focuses on an i-voting application, the FDKG concept is a general cryptographic primitive. Its applicability extends to any setting requiring robust, threshold-based secret reconstruction where global membership is uncertain or dynamic. Potential areas include collaborative data encryption/decryption in ad-hoc networks, decentralized credential management systems, and key management for emergent governance structures like Decentralized Autonomous Organizations (DAOs). Exploring these diverse applications and tailoring FDKG to their specific needs presents a rich avenue for future work.

\section{Discussion and Conclusions}
\label{sec:discussion-conclusion}

This paper introduced Federated Distributed Key Generation (FDKG), a novel protocol that enhances the flexibility and resilience of distributed key establishment in dynamic environments. Unlike standard DKG protocols that assume a fixed set of participants and a global threshold, FDKG empowers individual participants to select their own "guardian sets" of size $k$ for share recovery, accommodating scenarios with unknown or fluctuating network membership. Our security analysis (Section~\ref{sec:security_analysis}) demonstrates that FDKG achieves correctness and privacy for the key generation phase, and liveness for key reconstruction, provided an adversary does not control both a participant and $k-t+1$ of its chosen guardians during reconstruction.

FDKG generalizes traditional DKG by integrating principles of federated trust, reminiscent of Federated Byzantine Agreement (FBA) systems like the Stellar Consensus Protocol. This allows for dynamic trust delegation at the participant level. By achieving its core functionality in two communication rounds (generation and reconstruction), FDKG is suitable for systems involving human interaction, such as the i-voting application presented. The reliance on NIZK proofs (e.g., zk-SNARKs) for verifiability ensures protocol integrity but introduces computational overhead and, in the case of Groth16, a trusted setup requirement, which are important considerations for deployment. Future work could explore NIZKs with transparent setups (e.g., STARKs or Bulletproofs, though the latter might impact succinctness) to alleviate this.

Our simulations (Section~\ref{sec:liveness_simulations}) provided practical insights into FDKG's operational robustness. Liveness was shown to be heavily influenced by participant engagement (participation rate $p$) and continued availability (retention rate $r$). For example, in a network of $n=100$ parties with $p=0.8$ and $r=0.9$, FDKG consistently achieved liveness for thresholds $t < 0.7k$. However, with a lower retention of $r=0.5$, perfect liveness was only achieved for smaller thresholds $t \le 0.4k$ (Figure~\ref{fig:guardian_configs}). The choice of guardian selection topology also matters, with Barabási-Albert (BA) providing a slight resilience advantage over random selection in smaller networks due to its hub-centric nature (Figure~\ref{fig:network_model}), although this advantage diminishes for $n > 1000$.

Performance evaluations (Section~\ref{sec:performance_evaluation}) quantified the computational and communication costs. For $n=100$ participants with $k=40$ guardians, the FDKG Generation phase incurred approximately 336 kB of broadcast data per participant. NIZK proof generation, while enabling verifiability, is the main computational bottleneck, with $\pi_{FDKG_i}$ proving times ranging from $3.36$s (for $k=10$) to $14.79$s (for $k=100$) in our experiments (Table~\ref{table:perf_summary}). These figures highlight the trade-off between the flexibility offered by larger guardian sets and the associated costs, underscoring the need for careful parameter tuning based on specific application constraints: smaller $(k,t)$ values may be preferable for latency-sensitive systems, while larger values enhance resilience in more adversarial settings. A critical aspect of FDKG is balancing guardian set size ($k$) against security; while smaller sets reduce overhead, they concentrate trust, making individual guardians more critical for liveness if the original participant is unavailable or malicious.

We envision FDKG as a versatile cryptographic primitive for multi-party systems that require robust and verifiable key generation but must operate under conditions of incomplete or dynamic participation. Future research directions include optimizing NIZK proofs for FDKG-specific relations, refining vote decoding mechanisms for FDKG-based voting schemes, and developing more sophisticated, context-aware guardian selection strategies to further enhance FDKG's practical usability and security. By embracing a decentralized and adaptive ethos, FDKG offers a pathway to building scalable and secure distributed systems for diverse collaborative environments.

\section*{Reproducible Research}
To ensure the reproducibility of our research, all source code, datasets, and scripts used to generate results are publicly available. The repository is hosted at \url{https://github.com/stanbar/fdkg}.

\section*{Acknowledgments}
We thank Lev Soukhanov, 
Maya Dotan, Jonathan Katz, Oskar Goldhahn, and Thomas Haines for their insightful comments, enhancing the paper’s structure and cryptographic rigor.

The research was supported partially by the
project “Cloud Artificial Intelligence Service Engineering (CAISE)
platform to create universal and smart services for various application areas”, No. KPOD.05.10-IW.10-0005/24, as part of the European
IPCEI-CIS program, financed by NRRP (National Recovery and
Resilience Plan).

\bibliography{main}
\appendix

\section{Ideal Functionality and Proof Sketch}
\label{app:ideal_functionality}

\subsection{Ideal Functionality $\mathcal{F}_{FDKG}$}

The ideal functionality $\mathcal{F}_{FDKG}$ interacts with the parties $\Party{1}, \dots, \Party{n}$ and a simulator $\mathcal{S}$. It is parameterized by system parameters (group $\mathbb{G}$, generator $G$), FDKG parameters $(t, k)$, and dynamically tracks participation. Its formal description is provided in Figure~\ref{fig:ideal_functionality_fdkg}.

\begin{figure*}[!ht] 
\hrule\medskip
\centerline{\textbf{Ideal Functionality} $\mathcal{F}^{n,t,k}_{FDKG}$}
\medskip\hrule
\newcounter{fstep-fdkg-fig} 
\begin{list}{\arabic{fstep-fdkg-fig}.}{\usecounter{fstep-fdkg-fig}\leftmargin=1.5em\itemsep=2pt\labelwidth=1em\leftmargin=\labelwidth\labelsep=0.5em}
    
    \item \textbf{Initialization:}
        \begin{itemize}
            \item Waits to receive (Init, $\mathcal{C}$) from the simulator $\mathcal{S}$. Stores the set of corrupted parties $\mathcal{C}$. Let $\mathcal{H} = \Parties \setminus \mathcal{C}$.
            \item Initializes an empty set of participants $\SetOfFDKG = \emptyset$.
            \item Initializes empty storage for polynomials $f_i$ and guardian sets $\GuardianSetOf{i}$ for all $i \in \Parties$.
            \item Initializes state flag `phase` = 'activation'.
        \end{itemize}

    \item \textbf{Honest Party Activation:} Upon receiving (ActivateHonest, $i, \GuardianSetOf{i}$) from the simulator $\mathcal{S}$ (indicating honest party $\Party{i} \in \mathcal{H}$ decides to participate):
        \begin{itemize}
            \item If `phase` is 'activation' and $\Party{i} \notin \SetOfFDKG$:
                \begin{itemize}
                    \item Add $\Party{i}$ to $\SetOfFDKG$. 
                    \item Store the provided guardian set $\GuardianSetOf{i}$ (verify $|\GuardianSetOf{i}| = k$, $\Party{i} \notin \GuardianSetOf{i}$).
                    \item Choose a random polynomial $f_i(X)$ of degree $t-1$ over $\mathbb{Z}_q$ and store it.
                    \item Send (Activated, $i$) back to the simulator $\mathcal{S}$. 
                \end{itemize}
        \end{itemize}
        
    \item \textbf{Corrupt Party Activation:} Upon receiving (ActivateCorrupt, $i, f_i, \GuardianSetOf{i}$) from the simulator $\mathcal{S}$ (on behalf of $\Party{i} \in \mathcal{C}$):
        \begin{itemize}
             \item If `phase` is 'activation' and $\Party{i} \notin \SetOfFDKG$:
                 \begin{itemize}
                     \item Add $\Party{i}$ to $\SetOfFDKG$. 
                     \item Verify degree of $f_i$ is $t-1$, $|\GuardianSetOf{i}|=k$, $\Party{i} \notin \GuardianSetOf{i}$. Store valid $f_i$ and $\GuardianSetOf{i}$.
                 \end{itemize}
        \end{itemize}

    \item \textbf{Key Computation Trigger:} Upon receiving (ComputeKeys) from the simulator $\mathcal{S}$\footnote{Note on Round Termination: For simplicity in the ideal functionality $\mathcal{F}_{FDKG}$, we model the transition from the activation phase to the key computation phase via this `(ComputeKeys)` trigger sent by the simulator. In a real-world deployment, particularly on a blockchain, the end of the Generation phase would typically be determined by external environmental factors like block height or a predetermined time duration, not controlled by the adversary. Participants submitting messages after this cutoff would be excluded. Our security analysis focuses on the correctness and privacy properties based on valid participation within the allowed timeframe, assuming the round termination mechanism itself is robust. The core security guarantees derived from the underlying cryptography (IND-CPA PKE, NIZK soundness/ZK) are not fundamentally affected by this modeling simplification.}:
        \begin{itemize}
            \item If `phase` is 'activation':
                \begin{itemize}
                    \item Set `phase` = 'computed'.
                    \item If $\SetOfFDKG = \emptyset$: Send (KeysComputed, null, $\emptyset$, $\emptyset$, $\emptyset$, $\emptyset$) to $\mathcal{S}$ and stop this trigger processing.
                    \item For each $\Party{k} \in \SetOfFDKG$: Define $\PartialSecretKey{k} := f_k(0)$ and $\PartialPublicKey{k} := G^{\PartialSecretKey{k}}$. Let $\mathbb{E} = \{\PartialPublicKey{k}\}_{k \in \SetOfFDKG}$.
                    \item Compute global secret key $\SecretKey := \sum_{k \in \SetOfFDKG} \PartialSecretKey{k} \pmod{q}$ and global public key $\PublicKey := G^{\SecretKey}$.
                    \item For each $k \in \SetOfFDKG$ and each $j$ s.t. $\Party{j} \in \GuardianSetOf{k}$: Compute share $\SharePartialSecretKey{k}{j} := f_k(j)$.
                    \item Send (KeysComputed, $\PublicKey$, $\mathbb{E}$, $\{\GuardianSetOf{k}\}_{k \in \SetOfFDKG \cap \mathcal{H}}$, $\{(\PartialSecretKey{k}, f_k, \GuardianSetOf{k})\}_{k \in \SetOfFDKG \cap \mathcal{C}}$, $\{ \SharePartialSecretKey{k}{j} \mid k \in \SetOfFDKG, \Party{j} \in \GuardianSetOf{k} \cap \mathcal{C} \}$) to the simulator $\mathcal{S}$.
                    \item For each honest participant $\Party{i} \in \SetOfFDKG \cap \mathcal{H}$: Send (Output, $\PublicKey$, $\mathbb{E}$, $\PartialSecretKey{i}$, $\GuardianSetOf{i}$, $\{ \SharePartialSecretKey{k}{i} \mid k \in \SetOfFDKG, \Party{i} \in \GuardianSetOf{k} \}$) to $\Party{i}$.
                \end{itemize}
        \end{itemize}

\end{list}
\medskip\hrule
\caption{Ideal functionality for Federated Distributed Key Generation $\mathcal{F}^{n,t,k}_{FDKG}$.}
\label{fig:ideal_functionality_fdkg} 
\end{figure*}

\subsection{Proof Sketch for Theorem~\ref{thm:security_fdkg_generation_formal}}
\label{app:proof_sketch}

We show that $\Pi_{FDKG}$ securely realizes $\mathcal{F}_{FDKG}$ against any PPT
adversary $\mathcal{A}$ corrupting a static set $\Corrupted$. The proof proceeds
by a sequence of hybrids changing the honest parties' Round~1 messages.

\paragraph{Notation}
Write $\mathsf{View}_r$ for $\mathcal{A}$'s view in world $r$. All probabilities
are over the joint coin tosses of honest parties, the adversary, and the NIZK/PKE.

\medskip
\noindent\textbf{Hybrid H$_0$ (Real world).}
Honest parties follow $\Pi_{FDKG}$: they publish $(\PartialPublicKey{i},
\GuardianSetOf{i}, \{\EncryptedSharePartialSecretKey{i}{j}\}_{j\in \GuardianSetOf{i}}, \pi_{FDKG_i})$
where each $\EncryptedSharePartialSecretKey{i}{j}$ encrypts the \emph{real} share
$f_i(j)$ and $\pi_{FDKG_i}$ is a real proof.

\noindent\textbf{Hybrid H$_1$ (Simulated NIZKs).}
Same as H$_0$, except every honest $\pi_{FDKG_i}$ is replaced by a simulated proof
produced by the NIZK simulator for the same public statement.

\begin{lemma}[Zero-knowledge step]
\label{lem:zk}
$\mathsf{View}_{\text{H}_0} \approx_c \mathsf{View}_{\text{H}_1}$ under NIZK
zero-knowledge.
\end{lemma}

\noindent\textbf{Hybrid H$_2$ (IND-CPA switch for honest recipients).}
Same as H$_1$, except that for every honest dealer $i$ and honest recipient
$j\notin\Corrupted$, the ciphertext $\EncryptedSharePartialSecretKey{i}{j}$
is changed from an encryption of $f_i(j)$ to an encryption of $0$ under
$\textrm{pk}_j$. For corrupted recipients $j\in\Corrupted$, honest dealers still
encrypt the \emph{actual} share $f_i(j)$.

\begin{lemma}[IND-CPA step]
\label{lem:indcpa}
$\mathsf{View}_{\text{H}_1} \approx_c \mathsf{View}_{\text{H}_2}$ under PKE IND-CPA.
\end{lemma}

\noindent\textbf{Hybrid H$_3$ (Ideal world distribution).}
This is the ideal execution with simulator $\mathcal{S}$ interacting with
$\mathcal{F}_{FDKG}$:
\begin{itemize}
  \item For corrupt dealers, the simulator relays their real broadcasts.
  \item For each honest dealer $i$, the simulator uses the public $\PartialPublicKey{i}$
        and $\GuardianSetOf{i}$ supplied by $\mathcal{F}_{FDKG}$, encrypts $0$ for
        honest recipients, and encrypts the \emph{actual} shares $f_i(j)$ (obtained
        from $\mathcal{F}_{FDKG}$) for corrupted recipients $j\in\Corrupted$.
        Proofs are simulated.
\end{itemize}
By construction, $\mathsf{View}_{\text{H}_2}$ and $\mathsf{View}_{\text{H}_3}$
are \emph{identical}.

\begin{lemma}[Identity step]
\label{lem:identity}
$\mathsf{View}_{\text{H}_2} \equiv \mathsf{View}_{\text{H}_3}$.
\end{lemma}

\paragraph{Indistinguishability conclusion}
By Lemmas~\ref{lem:zk}–\ref{lem:identity} and a triangle inequality,
$\mathsf{View}_{\text{H}_0} \approx_c \mathsf{View}_{\text{H}_3}$, so the real
execution is computationally indistinguishable from the ideal one.

\paragraph{Correctness and Robustness from soundness}
In any world where honest broadcasts are \emph{accepted}, NIZK \emph{soundness}
implies that each accepted tuple of an (honest or corrupt) dealer $i$ is
consistent with a single degree-$(t-1)$ polynomial $f_i$ and with
$\PartialPublicKey{i}=G^{f_i(0)}$; otherwise the proof would be rejected except
with negligible probability. Hence each accepted ciphertext encrypts $f_i(j)$,
and the published $\PartialPublicKey{i}$ equals $G^{\PartialSecretKey{i}}$ with
$\PartialSecretKey{i}=f_i(0)$, yielding
$\PublicKey=\prod_{i\in\SetOfFDKG}\PartialPublicKey{i}=G^{\sum_{i\in\SetOfFDKG}\PartialSecretKey{i}}$.
Moreover, verification rejects malformed or equivocal broadcasts, so corrupted
parties cannot force acceptance of inconsistent data; their only effect is to
omit or abort. This establishes \emph{Correctness} and \emph{Robustness} of
Generation as stated in Theorem~\ref{thm:security_fdkg_generation_formal}.

\paragraph{Privacy of Generation}
In H$_2$ (hence in H$_3$), the adversary’s view on honest-to-honest ciphertexts
is distributed as encryptions of $0$; by Lemma~\ref{lem:indcpa}, this is
indistinguishable from encryptions of the true shares. Ciphertexts to corrupted
recipients contain the \emph{real} shares, matching the ideal leakage
$\{\SharePartialSecretKey{i}{j}\mid j\in\GuardianSetOf{i}\cap\Corrupted\}$.
Simulated proofs are indistinguishable by Lemma~\ref{lem:zk}. Therefore the view
leaks no additional information about honest $\{\PartialSecretKey{i}\}$ beyond
public values and the intended shares to corrupted recipients, establishing
\emph{Privacy} for Generation.

\section{zkSNARK Constructions}
\label{app:proofs}

This section details the specific statements and witnesses for the NIZK proofs used in the FDKG protocol and the associated voting application. We assume a NIZK proof system $(\textrm{Setup}, \textrm{Prove}, \textrm{Verify})$ operating with a common reference string $\textrm{crs}$ over a suitable group $\mathbb{G}$ with generator $G$ and order $q$. The proofs are instantiated using Groth16 \cite{grothSizePairingBasedNoninteractive2016}. Each subsection defines the relation $\mathcal{R}$ implicitly through the statement $x$ and witness $w$. Proofs are generated as $\pi \leftarrow \textrm{Prove}(\textrm{crs}, x, w)$ and verified by checking $\textrm{Verify}(\textrm{crs}, x, \pi) \stackrel{?}{=} 1$.

The detailed Circom circuit implementations for these proofs are publicly available on \url{https://github.com/stanbar/FDKG}.

\subsection{FDKG Proof ($\pi_{FDKG_i}$)}
\label{app:proof-fdkg}
Proves correct generation and encryption of shares by participant $\Party{i}$ during the Generation phase.
\begin{itemize}
    \item \textbf{Statement ($x_i$):} Consists of the party's partial public key $\PartialPublicKey{i}$, the set of PKE public keys $\{\textrm{pk}_j\}_{\Party{j} \in \GuardianSetOf{i}}$ for its guardians, and the set of corresponding encrypted shares $\mathbb{C}_i = \{\EncryptedSharePartialSecretKey{i}{j}\}_{\Party{j} \in \GuardianSetOf{i}}$.
    \item \textbf{Witness ($w_i$):} Consists of the coefficients $\{a_0, \dots, a_{t-1}\}$ of $\Party{i}$'s secret polynomial $f_i(X)$ (where $\PartialSecretKey{i} = a_0 = f_i(0)$), and the PKE randomness $\{(k_{i,j}, r_{i,j})\}_{\Party{j} \in \GuardianSetOf{i}}$ used to produce each ciphertext in $\mathbb{C}_i$.
    \item \textbf{Relation $\mathcal{R}_{FDKG}$ ($((x_i), w_i) \in \mathcal{R}_{FDKG}$):} The circuit (PVSS.circom) checks that:
        \begin{enumerate}
            \item The public $\PartialPublicKey{i}$ is correctly derived from the witness coefficient $a_0$ (i.e., $\PartialPublicKey{i} = G^{a_0}$).
            \item For each guardian $\Party{j} \in \GuardianSetOf{i}$:
                \begin{itemize}
                    \item The plaintext share $s_{i,j} = f_i(j)$ is correctly evaluated from the witness polynomial coefficients $\{a_k\}$.
                    \item The ciphertext $\EncryptedSharePartialSecretKey{i}{j} \in \mathbb{C}_i$ is a valid ElGamal variant encryption of the plaintext share $s_{i,j}$ under guardian $\Party{j}$'s PKE public key $\textrm{pk}_j$, using the witness randomness $(k_{i,j}, r_{i,j})$. This involves verifying the structure $C_1 = G^{k_{i,j}}$, $M=G^{r_{i,j}}$, $C_2 = (\textrm{pk}_j)^{k_{i,j}} \cdot M$, and $\Delta = M.x - s_{i,j}$.
                \end{itemize}
        \end{enumerate}
    \item \textbf{Circuit Implementation:} \url{https://github.com/stanbar/FDKG/blob/main/circuits/circuits/pvss.circom}
\end{itemize}

\subsection{Partial Secret Proof ($\pi_{PS_i}$)}
\label{app:proof-ps}
Proves knowledge of the partial secret key $\PartialSecretKey{i}$ corresponding to the partial public key $\PartialPublicKey{i}$ during reconstruction.
\begin{itemize}
    \item \textbf{Statement ($x_i$):} The partial public key $\PartialPublicKey{i}$.
    \item \textbf{Witness ($w_i$):} The partial secret key $\PartialSecretKey{i}$.
    \item \textbf{Relation $\mathcal{R}_{PS}$ ($((x_i), w_i) \in \mathcal{R}_{PS}$):} Checks that $\PartialPublicKey{i} = G^{\PartialSecretKey{i}}$.
    \item \textbf{Circuit Implementation:} \url{https://github.com/stanbar/FDKG/blob/main/circuits/circuits/elGamal_c1.circom}
\end{itemize}

\subsection{Share Decryption Proof ($\pi_{SPS_{j,i}}$)}
\label{app:proof-sps}
Proves correct decryption of an encrypted share $\EncryptedSharePartialSecretKey{j}{i}$ by the recipient party $\Party{i}$ (guardian) using its PKE secret key $\textrm{sk}_i$.
\begin{itemize}
    \item \textbf{Statement ($x_{j,i}$):} Consists of the encrypted share $\EncryptedSharePartialSecretKey{j}{i}=(C_1, C_2, \Delta)$, the recipient's (guardian $\Party{i}$'s) PKE public key $\textrm{pk}_i$, and the claimed plaintext share $s_{j,i}$.
    \item \textbf{Witness ($w_{j,i}$):} The recipient's (guardian $\Party{i}$'s) PKE secret key $\textrm{sk}_i$.
    \item \textbf{Relation $\mathcal{R}_{SPS}$ ($((x_{j,i}), w_{j,i}) \in \mathcal{R}_{SPS}$):} The circuit (decrypt\_share.circom) checks that applying the ElGamal variant decryption steps using $\textrm{sk}_i$ to $(C_1, C_2, \Delta)$ yields the plaintext $s_{j,i}$. That is, it computes $M = C_2 \cdot (C_1^{\textrm{sk}_i})^{-1}$ and verifies $M.x - \Delta = s_{j,i} \pmod{q}$.
    \item \textbf{Circuit Implementation:} \url{https://github.com/stanbar/FDKG/blob/main/circuits/circuits/decrypt_share.circom}
\end{itemize}


\subsection{Ballot Proof ($\pi_{Ballot_i}$)}
\label{app:proof-ballot}
Proves a ballot $\Ballot{i}$ is a valid ElGamal encryption of an allowed vote $\Vote{i}$ under the global public key $\PublicKey$.
\begin{itemize}
    \item \textbf{Statement ($x_{B_i}$):} The global public key $\PublicKey$ and the ballot $\Ballot{i} = (\BallotA{i}, \BallotB{i})$.
    \item \textbf{Witness ($w_{B_i}$):} The blinding factor $\BlindingFactor{i} \in \mathbb{Z}_q$ and the encoded vote message $M_{vote}$.
    \item \textbf{Relation $\mathcal{R}_{Ballot}$ ($((x_{B_i}), w_{B_i}) \in \mathcal{R}_{Ballot}$):} The circuit (encrypt\_ballot.circom) checks that:
        \begin{enumerate}
            \item $\BallotA{i} = G^{\BlindingFactor{i}}$.
            \item $\BallotB{i} = \PublicKey^{\BlindingFactor{i}} \cdot G^{M_{vote}}$.
            \item The witness $M_{vote}$ corresponds to one of the allowed vote encodings (e.g., $M_{vote} \in \{2^0, \dots, 2^{(c-1)m}\}$).
        \end{enumerate}
    \item \textbf{Circuit Implementation:} \url{https://github.com/stanbar/FDKG/blob/main/circuits/circuits/encrypt_ballot.circom}
\end{itemize}

\subsection{Partial Decryption Proof ($\pi_{PD_i}$)}
\label{app:proof-pd}
Proves correct computation of the partial decryption $\PartialDecryptionFrom{i}$ by party $\Party{i}$.
\begin{itemize}
    \item \textbf{Statement ($x_{PD_i}$):} The aggregated ElGamal component $\TotalA$, the computed partial decryption $\PartialDecryptionFrom{i}$, and the party's partial public key $\PartialPublicKey{i}$.
    \item \textbf{Witness ($w_{PD_i}$):} The party's partial secret key $\PartialSecretKey{i}$.
    \item \textbf{Relation $\mathcal{R}_{PD}$ ($((x_{PD_i}), w_{PD_i}) \in \mathcal{R}_{PD}$):} The circuit (partial\_decryption.circom) checks that:
        \begin{enumerate}
            \item $\PartialPublicKey{i} = G^{\PartialSecretKey{i}}$.
            \item $\PartialDecryptionFrom{i} = (\TotalA)^{\PartialSecretKey{i}}$.
        \end{enumerate}
    \item \textbf{Circuit Implementation:} \url{https://github.com/stanbar/FDKG/blob/main/circuits/circuits/partial_decryption.circom}
\end{itemize}

\subsection{Partial Decryption Share Proof ($\pi_{PDS_{j,i}}$)}
\label{app:proof-pds}
Proves correct decryption of an SSS share $s_{j,i}$ (originally from $\Party{j}$, decrypted by guardian $\Party{i}$) and its application to computing $\Party{i}$'s share of the partial decryption of $\TotalA$.
\begin{itemize}
    \item \textbf{Statement ($x_{PDS_{j,i}}$):} The aggregated ElGamal component $\TotalA$, the computed share of partial decryption $\SharePartialDecryptionFromTo{j}{i}$, the encrypted SSS share $\EncryptedSharePartialSecretKey{j}{i}=(C_1, C_2, \Delta)$, and the decrypting guardian $\Party{i}$'s PKE public key $\textrm{pk}_i$.
    \item \textbf{Witness ($w_{PDS_{j,i}}$):} The decrypting guardian $\Party{i}$'s PKE secret key $\textrm{sk}_i$.
    \item \textbf{Relation $\mathcal{R}_{PDS}$ ($((x_{PDS_{j,i}}), w_{PDS_{j,i}}) \in \mathcal{R}_{PDS}$):} The circuit (partial\_decryption\_share.circom) first internally computes the plaintext SSS share $s_{j,i} = \textrm{Dec}_{\textrm{sk}_i}(\EncryptedSharePartialSecretKey{j}{i})$ (verifying the decryption step) and then checks that $\SharePartialDecryptionFromTo{j}{i} = (\TotalA)^{s_{j,i}}$.
    \item \textbf{Circuit Implementation:} \url{https://github.com/stanbar/FDKG/blob/main/circuits/circuits/partial_decryption_share.circom}
\end{itemize}

\end{document}